\documentclass[hidelinks,onefignum,onetabnum]{siamart220329}

\usepackage{amsmath}
\usepackage{comment}
\usepackage{cite}

\newcommand{\e}{\varepsilon}

\usepackage{lipsum}
\usepackage{amsfonts}
\usepackage{graphicx}
\usepackage{epstopdf}
\usepackage{algorithmic}
\ifpdf
  \DeclareGraphicsExtensions{.eps,.pdf,.png,.jpg}
\else
  \DeclareGraphicsExtensions{.eps}
\fi

\newsiamremark{remark}{Remark}
\newsiamremark{hypothesis}{Hypothesis}
\crefname{hypothesis}{Hypothesis}{Hypotheses}
\newsiamthm{claim}{Claim}

\headers{A phenotype-structured RD model of glioma growth}{F. Ballatore, X. Ruan, C. Giverso and T. Lorenzi}

\title{A phenotype-structured reaction-diffusion model of avascular glioma growth}

\author{Francesca Ballatore%
  \thanks{Laboratoire J. A. Dieudonné, Université Côte d'Azur, Nice, France (\email{francesca.ballatore@univ-cotedazur.fr})} 
  \and Xinran Ruan%
  \thanks{School of Mathematical Sciences, Capital Normal University, Beijing, China (\email{xinran.ruan@cnu.edu.cn})}%
  \and Chiara Giverso%
  \thanks{Dipartimento di Scienze Matematiche ``G.L. Lagrange'', Politecnico di Torino, Italy (\email{chiara.giverso@polito.it}, \email{tommaso.lorenzi@polito.it})}%
  \and Tommaso Lorenzi%
  \footnotemark[3]}

\usepackage{amsopn}

\ifpdf
\hypersetup{
	pdftitle={Title},
	pdfauthor={Francesca Ballatore, Xinran Ruan, Chiara Giverso, and Tommaso Lorenzi}
}
\fi

\usepackage{txfonts}
\usepackage{mathtools}

\begin{document}
	
\maketitle

\begin{abstract}
We consider a phenotype-structured reaction-diffusion model of avascular glioma growth. The model describes the interaction dynamics between tumour cells and oxygen, and takes into account anisotropic cell movement and oxygen diffusion related to structural anisotropy of the brain’s extracellular environment. In this model, phenotypic heterogeneity of tumour cells is captured by a continuous phenotype-structuring variable, the value of which evolves due to phenotypic changes. We first analyse a one-dimensional version of the model and formally show, through a Hopf-Cole transformation, that it admits, in appropriate asymptotic regimes, phenotypically heterogeneous travelling wave solutions, wherein the locally prevailing cell phenotype varies across the wave due to the presence of oxygen gradients. This provides a mathematical formalisation for the emergence of intratumour phenotypic heterogeneity driven by differences in oxygen availability across the tumour. We then report on the results of both 1D simulations, which corroborate the results of formal asymptotic analyses, and 2D simulations, which also demonstrate the impact of anisotropy in cell movement and oxygen diffusion on tumour growth and on the phenotypic composition of the tumour edge. These results are complemented with additional results of 3D simulations, which are carried out on the geometry of the brain by using a hybrid finite difference-finite element method and integrating patient-specific magnetic resonance imaging data with diffusion tensor imaging data.
\end{abstract}
	
	
	\section{Introduction}	
Reaction-diffusion (RD) models of glioma growth play a prominent role in the mathematical modelling of the progression and response to therapy of brain tumours~\cite{Alfonso:2017,Falco:2021,Hatzikirou:2005,Ocana:2024,Preul:2015}. Following the pioneering studies~\cite{Swanson:2000, Swanson:2002b, Swanson:2002a, Swanson:2003, Swanson:2004, Tracqui:1995, Woodward:1996}, models of this type of increasing levels of biological realism and mathematical sophistication have been developed over the years~\cite{Gerlee:2012, Hathout:2016, Jbabdi:2005, Kim:2017, Kim:2014, Kim:2009, Papadogiorgaki:2013, Pham:2012, Rockne:2009, Suveges:2021, Swanson:2011}.
 	
	Stripped to an essential form, RD models of glioma growth comprise a RD equation for the density of tumour cells, wherein the reaction term takes into account cell proliferation and death, while an anisotropic diffusion term models cell movement through anisotropic fibres in the brain. This equation can then be coupled with additional RD equations for the dynamics of other biotic and abiotic components of the tumour micro-environment. 
	
The majority of the existing RD models of glioma growth rely on the assumption that cells in the tumour share the same phenotypic characteristics, which do not evolve in time. As such, these models implicitly ignore the fact that gliomas, like tumours with other cells of origin, exhibit significant phenotypic heterogeneity, which results from dynamic adaptations driven by selective pressures exerted by different components of the tumour micro-environment~\cite{Fayzullin:2019, Gatenby:2020, Gillies:2012, Lloyd:2016, Marusyk:2012, Sun:2015}. 

In particular, similarly to other types of avascular tumours, central to phenotype heterogeneity in avascular gliomas are differences in micro-environment between the hypoxic tumour core, where oxygen levels are significantly depleted due to consumption and limited diffusion of oxygen, and the well-oxygenated tumour edge, where oxygen is supplied through diffusion from surrounding tissues~\cite{Alfarouk:2013, Casciari:1992, Gatenby:2007, Ibrahim:2017, Molavian:2009}. The former is predominantly populated by slow-proliferating cells displaying a primarily glycolytic phenotype, which enables them to thrive in hypoxic conditions. The latter mainly comprises cells which instead display a more oxidative phenotype, exhibiting lower levels of resistance to hypoxia and faster proliferation rates.

A possible way of incorporating intratumour phenotype heterogeneity, rooted in spatial variability in oxygen distribution, into models of tumour dynamics consists in introducing a continuous structuring variable which represents the cell phenotype~\cite{Lorenzi:2025,Perthame:2006}. This variable provides a simple aggregate representation of the energy metabolism of tumour cells and its value can be related, for instance, to the cell level of expression of hypoxia-inducible factors (e.g. HIF-1)~\cite{Giatromanolaki:2001, Lee:2004, Semenza:2003}. Such a modelling approach, previously employed, for instance, in~\cite{Celora:2023, Celora:2021, Chiari:2023, Cho:2017, Cho:2020, Fiandaca:2022, Fiandaca:2021, Lorenzi:2018, Lorz:2015, Villa:2021}, makes it possible to simultaneously describe the spatial and evolutionary dynamics of tumour cells and their adaptation to spatially heterogeneous oxygen levels. 

In this paper, we consider a phenotype-structured RD model of avascular glioma growth. The model describes the interaction dynamics between tumour cells and oxygen, and takes into account anisotropic cell movement and oxygen diffusion related to structural anisotropy of the brain’s extracellular environment. In this model, phenotypic heterogeneity of tumour cells is captured by a continuous phenotype-structuring variable, the value of which evolves due to phenotypic changes. The model is then formulated as a non-local RD equation for the local phenotype density of tumour cells coupled with a RD equation for the local oxygen concentration. Based on the observation that glucose levels in biological tissues are usually high enough not to represent a limiting factor for the proliferation of cells~\cite{Gatenby:2003,Gatenby:2004,Gravenmier:2018}, for the sake of simplicity, the dynamics of the glucose concentration are not incorporated in the model, as previously done, for instance, in~\cite{Ardavseva:2020,Villa:2021}. 

We first analyse a one-dimensional version of the model and formally show, through a Hopf-Cole transformation, that it admits, in appropriate asymptotic regimes, phenotypically heterogeneous travelling wave solutions, wherein the locally prevailing cell phenotype varies across the wave due to the presence of oxygen gradients. We then report on the results of both 1D simulations, which corroborate the results of formal asymptotic analyses, and 2D simulations, which also demonstrate the impact of anisotropy in cell movement and oxygen diffusion on tumour growth and on the phenotypic composition of the tumour edge. These results are complemented with additional results of 3D simulations, which are carried out on the geometry of the brain by integrating patient-specific magnetic resonance imaging data with diffusion tensor imaging data.

The RD system considered here can be regarded as a generalisation of the one presented in~\cite{Villa:2021}, wherein additional effects of anisotropy in cell movement and oxygen diffusion related to non-uniform alignment of the fibres composing the brain’s extracellular environment are incorporated. From the analytical point of view, a key novelty of the present work is that, in contrast to~\cite{Villa:2021}, where the long-time behaviour of the solutions to the model equations was studied, here phenotypically heterogeneous travelling wave solutions are investigated. From the numerical point of view, while in~\cite{Villa:2021} numerical simulations were carried out, through an explicit finite difference method, for the model posed on a square spatial domain, here the model is solved numerically by using a hybrid finite difference-finite element method and employing a semi-implicit time discretisation, which makes it possible to carry out numerical simulations of the model posed on a wide range of spatial domains and incorporating anisotropic diffusion.
	
The rest of the paper is organised as follows. In Section~\ref{mathematical_model}, we describe the mathematical model. In Section~\ref{formal_asymptotic_analysis}, through formal asymptotic analyses, we study phenotype-structured travelling wave solutions. In Section~\ref{numerical_simulations}, we present the main results of numerical simulations. In Section~\ref{discussion_and_research_perspectives}, we conclude with a summary of key findings and a discussion of future directions.	
	
\section{Mathematical model}
\label{mathematical_model} 
The dynamics of the density of tumour cells with phenotype $y \in [0,1]$ (i.e. the local phenotype density), $n(t, \boldsymbol{x}, y)$, and the oxygen concentration, $S(t,\boldsymbol{x})$, at position $\boldsymbol{x} \in \Omega$ and time $t \in \left[0, \infty \right)$ are governed by the following RD system
	\begin{equation}
		\label{modello_dim}
		\begin{cases}
			\partial_t n - {\rm div} \left({\bf D}_n(\boldsymbol{x}) \,  \nabla n \right) = R(y,\rho,S) \, n + \beta \, \partial_{yy}^2 n \, , \quad y \in (0,1) \, ,
			\\\\
			\displaystyle{\rho(t,\boldsymbol{x}) := \int_{0}^1 n(t,\boldsymbol{x},y) \, {\rm d}y}  \, ,
			\\\\
			\displaystyle{\partial_t S - {\rm div} \left({\bf D}_S(\boldsymbol{x}) \,  \nabla S \right) = - \nu S \int_{0}^1 r(y) \, n(t,\boldsymbol{x},y) \, {\rm d}y} \, ,
		\end{cases} 
		\quad \boldsymbol{x} \in \Omega,
	\end{equation}
where $\rho$ is the cell density. Here the spatial domain $\Omega \subset \mathbb{R}^d$, with $d \geq 1$, is a bounded and connected set with boundary $\partial \Omega$. The variable $y$ represents  the cell metabolic phenotype. Specifically, building on the modelling approach adopted in~\cite{Ardavseva:2020,Villa:2021}, we assume cells with phenotype $y=0$ to have a fully oxidative metabolism and produce energy via aerobic respiration only, while cells with phenotype $y=1$ express a fully glycolytic metabolism and produce energy through anaerobic glycolysis only. Moreover, cells with phenotypes $y \in (0,1)$ produce energy via aerobic respiration and anaerobic glycolysis, and smaller values of $y$ correlate with a more oxidative and less glycolytic metabolism. 

\subsection{Summary of the terms in the non-local RD equation~\eqref{modello_dim} for $n$}
The second term on the left-hand side of the non-local RD equation~\eqref{modello_dim} for $n$ models anisotropic random movement of cells, which is (for simplicity) described as a diffusion process with diffusivity modelled by the symmetric positive-definite tensor ${\bf D}_n(\boldsymbol{x})$, the form of which depends on the alignment of the extracellular fibres. Moreover, the second term on the right-hand side takes into account spontaneous, heritable phenotypic changes across tumour cells~\cite{Huang:2013}. These are modelled through a linear diffusion term with coefficient $\beta  \in \mathbb{R}^+$, which is linked to the rate of phenotypic changes~\cite{Chisholm:2016, Lorenzi:2025}. Finally, the first term on the right-hand side is a non-local reaction term that takes into account proliferation and death of tumour cells. In more detail, the function $R(y, \rho, S)$ is the net proliferation rate (i.e. the difference between the rate of proliferation and the rate of death) of cells with phenotype $y$ at time $t$, under the local environmental conditions at position $\boldsymbol{x}$, which are determined by the cell density, $\rho(t,\boldsymbol{x})$, and the oxygen concentration, $S(t,\boldsymbol{x})$. This function, which can be regarded as the fitness landscape of the tumour~\cite{Chisholm:2016, Lorenzi:2025}, is defined as  
	\begin{equation}
		\label{R}
	R(y,\rho,S) := \alpha \left(p(y,S) - \dfrac{\rho}{\rho_0} \right) \, , \quad p(y,S) := r(y)\frac{S}{S_0} + f(y).
	\end{equation} 
	In the definition~\eqref{R} of the function $R$: the function $p(y,S)$ takes into account the effect of proliferation of cells in the phenotypic state $y$ at position $\boldsymbol{x}$ and time $t$ under the oxygen concentration $S(t,\boldsymbol{x})$; the term $\dfrac{\rho}{\rho_0}$ incorporates the effect of density-dependent inhibition of growth (i.e. the cessation of growth at sufficiently high cell density)~\cite{Lieberman:1981}, with the parameter $\rho_0 \in \mathbb{R}^+$ being proportional to the local carrying capacity of the tumour; and the parameter $\alpha \in \mathbb{R}^+$ is a scaling factor with the dimension of an inverse time, which measures the intensity of phenotypic selection acting on tumour cells (i.e. larger values of $\alpha$ correlate with stronger phenotypic selection). This definition of the function $R$ can be regarded as a natural generalisation of a logistic term in the case where inhibition of growth acting at the level of the density $\rho$ occurs in the presence of both continuous phenotype structuring and environmental heterogeneity linked to variability in the spatial distribution of the nutrient (i.e. oxygen) fuelling population growth.
	
In the definition~\eqref{R} of the function $p$, the first term and the second term take into account cell proliferation fuelled by aerobic respiration and anaerobic glycolysis, respectively. The parameter $S_0 \in \mathbb{R}^+$ is linked to the tissue oxygen concentration in physiological conditions (i.e. when there are no tumour cells), and the functions $r(y)$ and $f(y)$ are smooth and bounded real functions that satisfy the following assumptions:
	\begin{equation}
		\label{r_piccolo}
		r(0) = \gamma \, , \quad r(1) = 0 \, , \quad \dfrac{{\rm d} r}{{\rm d} y} < 0 \text{ on } (0,1) \, , \quad \gamma \in \mathbb{R}^+ \, ,
	\end{equation}
and 	
	\begin{equation}
		\label{f}
		f(0) = 0 \, , \quad f(1) = \zeta \, , \quad \dfrac{{\rm d} f}{{\rm d} y} > 0 \text{ on } (0,1) \, , \quad \zeta \in \mathbb{R}^+ \, .
	\end{equation}

Where possible, we retain general functional forms for $r(y)$ and $f(y)$. However, where needed, building upon the definitions employed in~\cite{Ardavseva:2020,Villa:2021}, we consider the functions
\begin{equation}
\label{def:modparfun}
r(y):= \gamma \left(1-y^2\right) \, , \quad f(y):=\zeta \left[1 - (1 - y)^2\right]  \, ,
\end{equation}
which clearly satisfy the general assumptions~\eqref{r_piccolo} and~\eqref{f}.
Furthermore, in the following we will assume that $\zeta < \gamma$. This assumption, together with~\eqref{r_piccolo} and~\eqref{f}, ensures that the maximum of $r(y)$, $\gamma$, attained at $y=0$, is greater than the maximum of $f(y)$, $\zeta$, attained at $y=1$, which is consistent with biological evidence indicating the presence of a fitness cost associated with a less efficient glycolytic metabolism~\cite{Basanta:2008}. Similarly to~\cite{Ardavseva:2020}, the ratio $\gamma/\zeta$ can be regarded as a measure of the fitness cost of glycolytic metabolism. Note that, as similarly mentioned in~\cite{Villa:2021}, under definitions~\eqref{def:modparfun}, the function $p(y,S)$ can be rewritten as
	\begin{equation}	
\label{def:rfparab}
p(y,S) := r(y) \, \frac{S}{S_0}  + f(y) = a(S) - b(S) \Big(y - h(S) \Big)^2
	\end{equation}
	with
		\begin{equation}	
\label{def:abh}
h(S) := \dfrac{\zeta}{\zeta + \gamma \ \frac{S}{S_0} } \,, \quad a(S) := \gamma \ \frac{S}{S_0}  + \zeta \, h(S) \ , \quad b(S) := \zeta + \gamma \ \frac{S}{S_0}  \, ,
\end{equation}
thus leading to a quadratic fitness landscape of the type of the approximate fitness landscapes which can be inferred from experimental data through regression techniques~\cite{lande1983measurement,otwinowski2014inferring,stinchcombe2008estimating}. Note also that, in the fitness landscape defined via~\eqref{def:rfparab}, the maximum fitness, $a(S)$, the nonlinear selection gradient, $b(S)$, and the fittest phenotype, $h(S)$, are functions of the oxygen concentration, consistently with biological evidence for the fact that the environmental conditions the cells are exposed to depend on the oxygen availability.

\subsection{Summary of the terms in the RD equation~\eqref{modello_dim} for $S$}
The second term on the left-hand side of the RD equation~\eqref{modello_dim} for $S$ corresponds to anisotropic diffusion of the oxygen molecules, with diffusivity modelled by the symmetric positive-definite tensor ${\bf D}_S(\boldsymbol{x})$. The term on the right-hand side models oxygen consumption by cells whose proliferation is fuelled by aerobic respiration, and the parameter $\nu \in \mathbb{R}^+$ is linked to the rate at which oxygen is consumed by cells using aerobic respiration to produce the energy required for their proliferation. The form of this term relies on the assumption that faster cell proliferation fuelled by aerobic respiration may conceivably demand greater consumption of oxygen (i.e. cells with phenotypes represented by smaller values of the structuring variable $y$ consume more oxygen)~\cite{Ardavseva:2020,Villa:2021}.

\subsection{RD system for the nondimensionalised local phenotype density and oxygen concentration}
To nondimensionalise the dependent variables in the RD system~\eqref{modello_dim} complemented with the definitions~\eqref{R}, we divide the equation for $n$ by $\rho_0$ (i.e. the parameter linked to the local carrying capacity of the tumour) and the equation for $S$ by $S_0$ (i.e. the parameter linked to the tissue oxygen concentration in physiological conditions). In so doing, introducing the notation
	\begin{equation}
		\label{rescaling}
		\hat{n}=\dfrac{n}{\rho_0} \, , \quad \hat{\rho}=\dfrac{\rho}{\rho_0} \, , \quad \hat{S}=\dfrac{S}{S_0} \, , \quad \hat{\nu}=\rho_0\nu \, ,
	\end{equation}
we obtain the following system
	\begin{equation}
		\label{modello}
		\begin{cases}
			\partial_t \hat{n} - {\rm div} \left({\bf D}_n(\boldsymbol{x}) \,  \nabla \hat{n} \right) = \alpha \left(r(y)\hat{S} + f(y) - \hat{\rho} \right) \, \hat{n} + \beta \, \partial_{yy}^2 \hat{n} \, , \quad y \in (0,1) \, ,
			\\\\
			\displaystyle{\hat{\rho}(t,\boldsymbol{x}) := \int_{0}^1 \hat{n}(t,\boldsymbol{x},y) \, {\rm d}y}  \, ,
			\\\\
			\displaystyle{\partial_t \hat{S} - {\rm div} \left({\bf D}_S(\boldsymbol{x}) \,  \nabla \hat{S} \right) = - \hat{\nu} \hat{S} \int_{0}^1 r(y) \, \hat{n}(t,\boldsymbol{x},y) \, {\rm d}y} \, ,
		\end{cases} 
		\quad \boldsymbol{x} \in \Omega,
		\end{equation}	
subject to initial data such that
	\begin{equation}
		\label{eq:ICsn}
\hat{n}(0,\boldsymbol{x},y)=\hat{n}^0(\boldsymbol{x},y) \, \quad \hat{n}^0 \geq 0 \ \text{ on } \ \overline{\Omega} \times [0,1] \, , \quad 0\le \int_{0}^1 \hat{n}^0(\cdot,y) \, {\rm d}y \le 1 \ \text{ on } \ \overline{\Omega}
	\end{equation}
and
	\begin{equation}
		\label{eq:ICsS}
\hat{S}(0,\boldsymbol{x})=\hat{S}^0(\boldsymbol{x}) \,, \quad 0\le \hat{S}^0 \le1 \ \text{ on } \ \overline{\Omega} \, .
	\end{equation}
	
We complement the non-local RD equation~\eqref{modello} for $\hat{n}$ with zero-flux boundary conditions on $\partial \Omega$ and at the endpoints of the phenotype domain, i.e. we impose the following homogeneous Neumann boundary conditions
	\begin{equation}
	\label{eq:BCsnx}
		\nabla \hat{n} \cdot \boldsymbol{u}=0  \ \text{ on } \ \partial \Omega
	\end{equation}
and
	\begin{equation}
	\label{eq:BCsny}
\partial_y \hat{n}{\big|}_{y=0} =0 \, , \quad \partial_y \hat{n}{\big|}_{y=1} =0 \, ,
	\end{equation}
where $\boldsymbol{u}$ is the unit normal to $\partial \Omega$ that points outwards from $\Omega$. On the other hand, we complement the RD equation~\eqref{modello} for $\hat{S}$ with the following Dirichlet boundary conditions
	\begin{equation}
	\label{eq:BCsS}
	\hat{S} = 1 \ \text{ on } \ \partial \Omega,
	\end{equation}
which implicitly rely on the assumption that, being the boundary of the spatial domain sufficiently far from the tumour border, the oxygen concentration along it remains equal to the tissue oxygen concentration in physiological conditions.

	
\section{Main analytical results}
\label{formal_asymptotic_analysis}
In this section, we formally show that, in appropriate asymptotic regimes, the RD system~\eqref{modello} admits phenotype-structured travelling front solutions, wherein the locally prevailing cell phenotype varies across the front as a result of variability in the oxygen concentration. 
	
The formal analysis we carry out here is set within the framework of previous work on the study of travelling fronts and concentration phenomena in non-local RD equations modelling the spatial-evolutionary dynamics of phenotype-structured populations. In summary, concentration phenomena in a variant of the RD system~\eqref{modello} were studied in~\cite{jabin2023selection,mirrahimi2015asymptotic} taking ${\bf D}_n \equiv 0$ and making a quasi-stationary equilibrium assumption for the oxygen concentration. Moreover, focusing on non-local RD equations of the form of equation~\eqref{modello} for $\hat{n}$ but with isotropic diffusion: travelling fronts in the presence of reaction terms modelling shifting environments and spatially periodic environments were studied in \cite{alfaro2013travelling,alfaro2017effect} and \cite{boutillon2024reaction}, respectively; the long-time behaviour of the solution in the asymptotic regime of small diffusivity was studied in~\cite{bouin2015hamilton}; travelling fronts in the presence of constant proliferation rate and either constant or phenotype-dependent diffusivity were studied, respectively, in~\cite{benichou2012front,berestycki2015existence,bouin2012invasion} and~\cite{bouin2017super,bouin2014travelling,turanova2015model}; concentration phenomena at steady state, in the asymptotic regime of small rates of phenotypic changes, when the proliferation rate does not depend on the phenotype variable and there is phenotype-dependent diffusivity or both phenotype-dependent diffusivity and chemotaxis were studied in~\cite{lam2017integro} and~\cite{hao2019concentration}, respectively. Finally, both travelling fronts and steady-state solutions in a non-local RD equation of a form tangentially related to that of equation~\eqref{modello} for $\hat{n}$ but modelling the spatial-evolutionary dynamics of sexual populations were studied in~\cite{mirrahimi2013dynamics}. We refer the interested reader to~\cite{Lorenzi:2025} and references therein for a more comprehensive literature review on the subject.


	\subsection{Travelling wave problem}
	\label{modello_epsilon}
	First, for compactness of notation, we drop the carets in the system~\eqref{modello}. Then, to strip down the problem to its essence and facilitate  analysis of the model, we restrict our attention to a one-dimensional unbounded spatial domain, thereby setting $\boldsymbol{x} \equiv x \in \mathbb{R}$, and assume ${\bf D}_n(x) \equiv D_n \in \mathbb{R}^+$ and ${\bf D}_S(x) \equiv D_S  \in \mathbb{R}^+$. Next, to investigate the dynamics of the system for large $t$, we introduce a small parameter $\e \in \mathbb{R}^+$ and employ the time scaling $t \to t/\e$. Finally, we consider a scenario where oxygen diffusion and consumption occur on a slower time scale compared to cell proliferation and death, and cell movement and phenotypic changes occur on a slower time scale compared to oxygen diffusion and consumption. We thus employ the following parameter scaling:
	\begin{equation}
		\label{req:scaling}
D_S := \e \ , \quad \nu := \e \ , \quad D_n := \e^2 \ , \quad \beta := \e^2 \, .
	\end{equation}	
In this framework, the local phenotype density, $n_{\e}(t,x,y) \equiv n(\frac{t}{\e},x,y)$, and the oxygen concentration, $S_{\e}(t,x) \equiv S(\frac{t}{\e},x)$, satisfy the following rescaled RD system:
	\begin{equation}
		\label{rescaled_equation_2D}
		\begin{cases}
			\displaystyle{\e \, \partial_t n_{\e} - \e^2 \, \partial^2_{xx} n_{\e} = \alpha \, \Big(r(y) \, S_{\e}+f(y)-\rho_{\e} \Big) \, n_{\e} + \e^2 \partial_{yy}^2 n_{\e}} \, , \quad y \in (0,1) \, ,
			\\\\
			\displaystyle{\rho_{\e}(t,x) := \int_{0}^1 n_{\e}(t,x,y) \, {\rm d}y} \, ,
			\\\\
			\displaystyle{\partial_t S_{\e} - \partial^2_{xx} S_{\e}   = - S_{\e} \int_{0}^1 r(y) \, n_{\e}(t,x,y) \, {\rm d}y} \, ,
		\end{cases} 
		x \in \mathbb{R} \, .
	\end{equation}	
We complement the equation \eqref{rescaled_equation_2D} for $n_{\e}$ with homogeneous Neumann boundary conditions analogous to~\eqref{eq:BCsny}, that is, 
$$
\partial_y n_\e{\big|}_{y=0} =0 \, , \quad \partial_y n_\e{\big|}_{y=1} =0 \, .
$$
Note that, under these boundary conditions, integrating the equation \eqref{rescaled_equation_2D} for $n_{\e}$ with respect to $y$ gives
	\begin{equation}
		\label{rescaled_equation_2Drhoeps}
\e \, \partial_t \rho_{\e} - \e^2 \, \partial^2_{xx} \rho_{\e} = \alpha \, \int_{0}^1 \Big(r(y) \, S_{\e}+f(y)-\rho_{\e} \Big) \, n_{\e}\, {\rm d}y	\, , \quad x \in \mathbb{R} \, .
	\end{equation}

We seek travelling wave solutions of the system~\eqref{rescaled_equation_2D} with the $n_{\e}$-component that exhibits phenotype structuring, that is, solutions such that
$$
n_{\e}(t,x,y) \equiv n_{\e}(z,y) \, , \quad \rho_{\e}(t,x) \equiv \rho_{\e}(z) \, , \quad S_{\e}(t,x) \equiv S_{\e}(z) \, , \quad z = x-c \, t \, , \quad c \in \mathbb{R}^+ \, ,
$$
where $c$ is the wave speed. Substituting this travelling-wave ansatz into the system~\eqref{rescaled_equation_2D} and rearranging terms we obtain the following system:
	\begin{equation}	
			\label{rescaled_equation_2DTW}
		\begin{cases}
			\displaystyle{- \e^2 \, \partial^2_{zz} n_{\e} - \e \, c \, \partial_z n_{\e} = \alpha \, \Big(r(y)S_{\e}+f(y)-\rho_{\e} \Big) \, n_{\e} + \e^2 \partial_{yy}^2 n_{\e}} \, , \quad y \in (0,1)
			\\\\
			\displaystyle{\rho_{\e}(z) := \int_{0}^1 n_{\e}(z,y) \, {\rm d}y} \, ,
			\\\\
			\displaystyle{S''_{\e} + c \, S'_{\e} = S_{\e} \int_{0}^1 r(y) \, n_{\e}(z,y) \, {\rm d}y} \, ,
		\end{cases} 
				z \in \mathbb{R}.
	\end{equation}
Moreover, substituting the travelling-wave ansatz into equation~\eqref{rescaled_equation_2Drhoeps} and rearranging terms yields
	\begin{equation}
		\label{rescaled_equation_2DrhoepsTW}
\e^2 \, \rho''_{\e} + \e \, c\, \rho'_{\e} + \alpha \, \int_{0}^1 \Big(r(y) \, S_{\e}+f(y)-\rho_{\e} \Big) \, n_{\e}\, {\rm d}y = 0	\, , \quad z \in \mathbb{R} \, .
	\end{equation}	
	

	\subsection{Formal asymptotic analysis} Building on previous studies of concentration phenomena in phenotype-structured partial differential equations, see the seminal papers~\cite{barles2009concentration,Diekmann:2005,lorz2011dirac,perthame2008dirac}, and phenotypic structuring across travelling waves in space- and phenotype-structured partial differential equations, see the recent review~\cite{Lorenzi:2025} and references therein, we make the following Hopf-Cole transformation~\cite{Barles:1990,Evans:1989,Fleming:1986}:
		\begin{equation}
		\label{ansatz}
		n_{\e}(z, y)=e^{\frac{u_{\e}(z, y)}{\e}} \, .
	\end{equation}
		Moreover, we denote by $\rho(z)$, $u(z,y)$, and $S(z)$ the leading-order terms of the asymptotic expansions for $\rho_{\e}(z)$, $u_{\e}(z,y)$, and $S_{\e}(z)$ as $\e \to 0$. 

\subsubsection{Concentration phenomena along travelling waves} Considering $z \in \mathbb{R}$ such that $\rho(z)>0$, i.e. $z \in \operatorname{Supp}(\rho)$, substituting the ansatz~\eqref{ansatz} into the equation~\eqref{rescaled_equation_2DTW} for $n_{\e}$ and letting $\e \to 0$ we formally obtain the following Hamilton-Jacobi equation for $u(z, y)$:
	\begin{equation}
	\label{travelling1A}
		-c \, \partial_z u - \left(\partial_{z}u \right)^2 = \alpha\Big(r(y)S+f(y)-\rho(z)\Big) + \left(\partial_y u\right)^2 \, , \quad \ (z, y) \in \operatorname{Supp}(\rho) \times (0, 1) \, .
		\end{equation}	
		
When $\rho_{\e}(z) < \infty$ for all $\e >0$ and $\rho(z) < \infty$ for all $z \in \mathbb{R}$, if $u_{\e}(z, y)$ is a strictly concave function of $y$, with a unique non-degenerate maximum point at $\bar{y}_{\e}(z)$, and $u(z, y)$ is also a strictly concave function of $y$, with a unique non-degenerate maximum point at $\bar{y}(z)$, letting $\e \to 0$ in \eqref{ansatz} formally gives the constraint 
	\begin{equation}
		\label{constraint_2A}
		u(z, \bar{y}(z))=  \max _{y \in[0, 1]} u(z, y) = 0 \, , \quad z \in \operatorname{Supp}(\rho) \, .
	\end{equation}
In fact, if it was $u(z, \bar{y}(z))>0$ then $\rho_{\e}(z) \to \infty$ as $\e  \to 0$ and thus it could not be that $\rho(z) < \infty$, whereas if it was $u(z, \bar{y}(z))<0$ then $\rho_{\e}(z) \to 0$ as $\e  \to 0$ and thus it could not be that $\rho(z)>0$. 

Moreover, from~\eqref{constraint_2A}, since $\displaystyle{u(z, \bar{y}(z))=  \max _{y \in[0, 1]} u(z, y)}$ and $u(z, \bar{y}(z))=0$ for all $z \in \operatorname{Supp}(\rho)$, it formally follows that 
\begin{equation}
		\label{condition_x_and_t}
		\partial_y u(z,\bar{y}(z)) = 0  \quad \text{and} \quad \partial_z u(z,\bar{y}(z)) = 0 \, , \quad z \in \operatorname{Supp}(\rho) \ .
	\end{equation}
	
Finally, the constraint~\eqref{constraint_2A} also implies that, in the asymptotic regime $\e \to 0$, $n_{\e}(z,y)$ concentrates on the singleton set $\{\bar{y}(z)\}$ for all $z \in \operatorname{Supp}(\rho)$. This along with the fact that $\rho_{\e}(z) := \int_{0}^1 n_{\e}(z,y) \, {\rm d}y$ converges to $\rho(z)$ as $\e \to 0$ formally give 
\begin{equation}
\label{eq:convtodirac}
n_{\varepsilon}(z,y) \xrightharpoonup[\varepsilon \rightarrow 0]{} \rho(z) \, \delta_{\bar{y}(z)}(y) \quad \text{weakly in measures},
\end{equation} 
where $\delta_{\bar{y}(z)}(y)$ is the Dirac delta centred at $y=\bar{y}(z)$.
The result~\eqref{eq:convtodirac} indicates that concentration phenomena emerge for $\varepsilon \rightarrow 0$, i.e. $n_{\varepsilon}(z,y)$ becomes concentrated as a weighted Dirac mass along the $y-$dimension. The concentration point $y=\bar{y}(z)$ (i.e. the centre of the Dirac mass) can thus be biologically interpreted as the phenotype that is expressed by tumour cells at position $z$ along the front (i.e. the locally prevailing cell phenotype), and the weight $\rho(z)$ is the cell density at position $z$ along the front.
	
\subsubsection{Expression of  $\rho(z)$ and differential equations for $S(z)$ and $\bar{y}(z)$} 				
Letting $\e \to 0$ in equation~\eqref{rescaled_equation_2DrhoepsTW}, recalling~\eqref{eq:convtodirac}, we formally find that, for any $z \in \mathbb{R}$, $\rho(z)$ attains either the value $0$ or the value $r\left(\bar{y}(z)\right)S(z)+f\left(\bar{y}(z)\right)$, as it can also be seen by evaluating \eqref{travelling1A} at $\bar{y}(z)$ and then using \eqref{constraint_2A} and \eqref{condition_x_and_t}. We thus consider travelling fronts such that, up to translation, 
				\begin{equation}	
			\label{eq:shapeofrho}
\rho(z) = 
\begin{cases}
r\left(\bar{y}(z)\right)S(z)+f\left(\bar{y}(z)\right)  \, , \quad z < 0
\\
0  \, , \qquad\qquad\qquad\qquad\quad\; z \geq 0 \, ,
\end{cases}
					\end{equation}	
with $z=0$ being the location of the tumour border. Hence, $\operatorname{Supp}(\rho)=(-\infty,0)$.

Moreover, letting $\varepsilon \to 0$ in the differential equation~\eqref{rescaled_equation_2DTW} for $S_{\e}(z)$ and using~\eqref{eq:convtodirac}, we formally obtain the following differential equation for $S$:
\begin{equation}
\label{eqSTW}
S''(z) + c \, S'(z) = S(z) \, r(\bar{y}(z)) \, \rho(z) \, , \quad z \in \mathbb{R} \, .
\end{equation} 

Furthermore, differentiating~\eqref{travelling1A} with respect to $y$ yields
$$
-c \, \partial^2_{zy} u - 2 \, \partial_{z}u \, \partial^2_{zy}u = \alpha \left(\frac{{\rm d} r(y)}{{\rm d}y}S+\frac{{\rm d} f(y)}{{\rm d}y}\right)+2 \, \partial_y u \, \partial^2_{yy} u \, , \quad \ (z, y) \in \operatorname{Supp}(\rho) \times (0, 1) \, .
$$
Evaluating the above equation at $\bar{y}(z)$ and using conditions~\eqref{condition_x_and_t} we obtain
	\begin{equation}
		\label{d_ty_2A}
		- \partial^2_{zy} u(z, \bar{y}(z)) = \dfrac{\alpha}{c} \, \left(\frac{{\rm d} r}{{\rm d}y}( \bar{y}(z))S(z)+\frac{{\rm d} f}{{\rm d}y}( \bar{y}(z))\right) \, , \quad \ z \in \operatorname{Supp}(\rho) \, .
	\end{equation}
Then, differentiating the first condition of \eqref{condition_x_and_t} with respect to $z$, and using the fact that $\partial_{yy}^2 u(z, \bar{y}(z)) < 0$, since $\bar{y}(z)$ is the unique non-degenerate  maximum point of $u(z,y)$, formally gives 
$$
\partial_{zy}^2 u(z, \bar{y}(z))+\partial_{yy}^2 u(z, \bar{y}(z)) \, \bar{y}'(z)=0 \; \Rightarrow \; \, \bar{y}'(z)  = \dfrac{-\partial_{zy}^2 u(z, \bar{y}(z))}{ \partial_{yy}^2 u(z, \bar{y}(z))} , \quad z \in \operatorname{Supp}(\rho) \, .
$$
Finally, substituting the expression~\eqref{d_ty_2A} of $- \partial^2_{zy} u(z, \bar{y}(z))$ into the above equation leads to the following differential equation for $\bar{y}$:
	\begin{equation}
		\label{transport_equation_bar_y}
	\bar{y}'(z)  = \dfrac{\alpha}{c \, \partial_{yy}^2 u(z, \bar{y}(z))} \, \left(\frac{{\rm d} r}{{\rm d}y}( \bar{y}(z))S(z)+\frac{{\rm d} f}{{\rm d}y}( \bar{y}(z))\right) \, , \quad z \in \operatorname{Supp}(\rho) \, .		
	\end{equation}

\subsubsection{Boundary conditions} We consider travelling fronts corresponding to an invading tumour whereby: the hypoxic tumour core (i.e. where the rescaled oxygen concentration attains the minimum value $0$) is located at $z=-\infty$, while the region far ahead of the tumour (i.e. where the rescaled oxygen concentration attains the maximum value $1$) is located at $z=\infty$. Hence, we complement the differential equation~\eqref{eqSTW} with the following boundary conditions 
				\begin{equation}	
			\label{rescaled_equation_STWBCs}
			S(-\infty) = 0 \, , \quad S(\infty) = 1 \, .
					\end{equation}	
Specifically, we consider solutions of the differential equation~\eqref{eqSTW} subject to the boundary conditions~\eqref{rescaled_equation_STWBCs} which are such that
\begin{equation}
\label{propS2}
S'(-\infty) = 0 \quad \mbox{and} \quad S'(\infty) = 0 \, .
\end{equation}

Furthermore, since we expect the hypoxic tumour core at $z=-\infty$ to be composed of cells with phenotype $y=1$ (i.e. cells with a fully glycolytic metabolism), we complement the differential equation~\eqref{transport_equation_bar_y} with the following boundary condition
				\begin{equation}	
			\label{transport_equation_bar_yBCs}
			\bar{y}(-\infty) = 1 \, .
					\end{equation}
												
\subsubsection{Monotonicity of $S$ and formula for the wave speed} For $c \in \mathbb{R}^+$ fixed, in light of the nonnegativity of $r(y)$ and the expression~\eqref{eq:shapeofrho} of $\rho(z)$, the maximum principle ensures that the solution to the differential equation~\eqref{eqSTW}, subject to the boundary conditions~\eqref{rescaled_equation_STWBCs}, satisfying conditions~\eqref{propS2} is such that
\begin{equation}
\label{propS1}
0 < S(z) < 1 \; \mbox{ and } \; S'(z)  > 0 \; \; \mbox{ for } \; -\infty < z < \infty \, .
\end{equation}
Moreover, integrating the differential equation~\eqref{eqSTW} over $\mathbb{R}$, imposing the boundary conditions~\eqref{rescaled_equation_STWBCs}, and using~\eqref{propS2} we obtain
$$
0 + c \, \left(1-0 \right) = \int_{\mathbb{R}} S(z) \, r(\bar{y}(z)) \, \rho(z) \, {\rm d}z \, ,
$$
and then, substituting into the above equation the expression~\eqref{eq:shapeofrho} of $\rho(z)$, we find the following formula for the wave speed:
\begin{equation}
\label{wave_speed}
c = \int_{\operatorname{Supp}(\rho)} S(z) \, r(\bar{y}(z)) \, \Big[r\left(\bar{y}(z)\right)S(z)+f\left(\bar{y}(z)\right)\Big] \, {\rm d}z \, .
\end{equation}

\subsubsection{Shape of phenotype-structured travelling fronts} Substituting~\eqref{def:rfparab} into~\eqref{eq:shapeofrho} and \eqref{transport_equation_bar_y} yields, respectively, 
	\begin{equation}
		\label{rhospec1}
		\rho(z)= a(S(z)) - b(S(z)) \Big(\bar{y}(z) - h(S(z)) \Big)^2 \, , \quad z \in \operatorname{Supp}(\rho) 
	\end{equation}
and	
		\begin{equation}
		\label{transport_equation_bar_yspec1}
	\bar{y}'(z)  = -  \alpha \dfrac{2 \, b(S(z))}{c \,  \partial_{yy}^2 u(z, \bar{y}(z))} \, \Big(\bar{y}(z) - h(S(z)) \Big) \, , \quad z \in \operatorname{Supp}(\rho) \, .	
	\end{equation}

Letting $\alpha \to \infty$ in the differential equation~\eqref{transport_equation_bar_yspec1}, which corresponds to considering the asymptotic regime of strong phenotypic selection, and recalling the expression~\eqref{def:abh} of $h(S)$ formally gives 
		\begin{equation}
		\label{bar_yspec2}
  \bar{y}(z) = h(S(z)) = \dfrac{\zeta}{\zeta + \gamma \ S(z)} \, , \quad z \in \operatorname{Supp}(\rho)	 \, .
	\end{equation}
	In turn, substituting~\eqref{bar_yspec2} into~\eqref{rhospec1}, and recalling the expression~\eqref{def:abh} of $a(S)$, we find	
		\begin{equation}
		\label{rhospec2}
		\rho(z)= a(S(z)) = \gamma \ S(z) + \dfrac{\zeta^2}{\zeta + \gamma \ S(z)}  \, , \quad z \in \operatorname{Supp}(\rho) \, .
	\end{equation}
	
Under the boundary conditions~\eqref{rescaled_equation_STWBCs} and~\eqref{transport_equation_bar_yBCs}, the properties~\eqref{propS2} and~\eqref{propS1} of $S(z)$ along with the expression~\eqref{bar_yspec2} of $\bar{y}(z)$ and the expression~\eqref{rhospec2} of $\rho(z)$ allow us to conclude that travelling fronts that, up to translation, satisfy~\eqref{eq:shapeofrho} are such that 
		\begin{equation}
		\label{baryrhospec3}
\bar{y}'(z)  < 0 \; \mbox{ and } \; \rho'(z)  > 0 \; \; \mbox{ for } \; -\infty < z < 0 \, ,
	\end{equation}
\begin{equation}
\label{rho_bary_edge}
\rho(z) \to \gamma S(0) + \dfrac{\zeta^2}{\zeta + \gamma \, S(0)} , \quad \bar{y}(z) \to  \dfrac{\zeta}{\zeta+\gamma \, S(0)} , \quad \text{as} \ z \to 0^{-} \, ,
\end{equation}
and $\rho(z)$ jumps to $0$ at $z=0$.

 In summary,  the analytical results formally obtained in this section, which are confirmed by the results of numerical simulations presented in Figure~\ref{plot_rescaled}, as discussed in detail in Section~\ref{Rescaled_model}, indicate that: the oxygen concentration increases monotonically along the front, going from $S=0$ to $S=1$ (cf. \eqref{rescaled_equation_STWBCs} and \eqref{propS1}). The locally prevailing cell phenotype decreases monotonically along the front, connecting $\bar{y}=1$ to $\bar{y}=\frac{\zeta}{\zeta+\gamma \, S(0)} < 1$ (cf. \eqref{bar_yspec2}, \eqref{baryrhospec3}, and~\eqref{rho_bary_edge}). On the other hand, the cell density increases monotonically along the front, departing from $\rho=\zeta$ and approaching $\rho=\gamma \, S(0) + \frac{\zeta^2}{\zeta + \gamma \, S(0)} > \zeta$ (cf. \eqref{rhospec2}, \eqref{baryrhospec3} and~\eqref{rho_bary_edge}, and recall that $\gamma>\zeta$). 
 
Taken together, these results formalise the idea that differences in oxygenation across the tumour create distinct micro-environments, whereby the oxygenated edge and the hypoxic core of the tumour exhibit distinct phenotype compositions, with cells with a more oxidative metabolism being located around the edge of the tumour, while the regions in the proximity of the core are predominantly composed of cells with a primarily glycolytic metabolism. This promotes the emergence of intratumour phenotypic heterogeneity. Note also that the value of the locally prevailing cell phenotype, $\bar{y}$, at the edge of the front, that is, $\frac{\zeta}{\zeta+\gamma \, S(0)}$, approaches $0$ as the ratio $\frac{\gamma}{\zeta}$ increases. This, along with the monotonicity of $\bar{y}$ and the fact that at the back of the wave $\bar{y}=1$, indicates that the spectrum of values spanned by $\bar{y}$ widens in the presence of larger values of $\frac{\gamma}{\zeta}$. Since, as mentioned earlier, the ratio $\frac{\gamma}{\zeta}$ provides a measure of the fitness cost of glycolytic metabolism, these theoretical results indicate that higher fitness costs of glycolytic metabolism correlate with higher levels of intratumour phenotypic heterogeneity.

\section{Main numerical results}
	\label{numerical_simulations}
In this section, we present the main results of numerical simulations. In Section~\ref{numerical_method}, the scheme employed to solve numerically the initial-boundary value problem for the RD system of the model is described. In Section~\ref{Rescaled_model}, we compare the numerical solutions of the rescaled one-dimensional model~\eqref{rescaled_equation_2D} with the results of the formal asymptotic analyses carried out in Section~\ref{formal_asymptotic_analysis}. In Sections~\ref{Non_rescaled_model_in_a_square_domain} and~\ref{sec:brainsim} we present numerical solutions of the RD system~\eqref{modello} posed, respectively, on a square spatial domain and on the 3D-geometry of the brain.
	
\subsection{Numerical scheme}
\label{numerical_method}
We describe the numerical scheme employed for the simulations of the RD system~\eqref{modello} considered in Sections~\ref{Non_rescaled_model_in_a_square_domain} and~\ref{sec:brainsim}. This scheme is based on a hybrid finite difference-finite element method and relies on a time-splitting approach, which makes it possible to solve separately the conservative and non-conservative parts of the model equations. For the rescaled one-dimensional model~\eqref{rescaled_equation_2D} considered in Section~\ref{Rescaled_model}, a numerical scheme based on a purely finite difference method is used instead, which also relies on a time-splitting approach in order to split the conservative and non-conservative parts of the model equations.

\subsubsection{Time-splitting approach underlying the scheme}
Adopting a time-splitting approach, we decompose the RD equation~\eqref{modello} for $\hat{S}$ into the diffusion equation 
	\begin{equation}
	\label{eq:ns1}
	\partial_t \hat{S} - {\rm div} \left({\bf D}_S(\boldsymbol{x}) \,  \nabla \hat{S} \right) = 0 \, , \quad \boldsymbol{x} \in \Omega \, ,
	\end{equation}
subject to the boundary conditions~\eqref{eq:BCsS}, and the reaction equation 
	\begin{equation}
		\label{eq:ns2}
	\partial_t \hat{S} = - \hat{\nu} \, \hat{S} \int_{0}^{1} r(y) \, \hat{n}(t, \boldsymbol{x}, y) \, \mathrm{d}y \, , \quad \boldsymbol{x} \in \overline{\Omega} \ .
\end{equation}
These equations are sequentially solved at each time step. 

Similarly, for the non-local RD equation~\eqref{modello} for $\hat{n}$, at each time step we solve sequentially the diffusion equation 
\begin{equation}
	\label{eq:ns3}
	\partial_t \hat{n} - \operatorname{div} \left( \mathbf{D}_n(\boldsymbol{x}) \, \nabla \hat{n} \right) = 0 \, , \quad (\boldsymbol{x}, y) \in \Omega \times [0,1] \, ,
\end{equation}
subject to the boundary conditions~\eqref{eq:BCsnx}, then the diffusion equation 
	\begin{equation}
			\label{eq:ns4}
		\partial_t \hat{n} = \beta \, \partial_{yy}^2 \hat{n} \, , \quad (\boldsymbol{x},y) \in \overline{\Omega} \times (0,1) \ ,
	\end{equation}
subject to the boundary conditions~\eqref{eq:BCsny}	, and lastly the non-local reaction equation 
		\begin{equation}
		\label{eq:ns5a}
		\begin{cases}
		\partial_t \hat{n} = \alpha \left(r(y)\hat{S} + f(y) - \hat{\rho} \right) \, \hat{n} \, , \quad (\boldsymbol{x},y) \in \overline{\Omega} \times [0,1] 
		\\\\
	\displaystyle{\hat{\rho}(t,\boldsymbol{x}) := \int_{0}^1 \hat{n}(t,\boldsymbol{x},y) \,  {\rm d}y \ .}
		\end{cases}
	\end{equation}
In order to employ a semi-implicit time discretisation, as detailed in the next subsection, so as to ensure a balance between numerical stability and computational efficiency, instead of solving directly the non-local reaction equation~\eqref{eq:ns5a}, we make the change of variables $\hat{u} =  \ln(\hat{n})$, as similarly done in~\cite{Lorenzi:2021}. This change of variables is well-defined in the regions where $\hat{n} > 0$, which are the relevant regions, since where $\hat{n} = 0$ the right-hand side of the non-local reaction equation~\eqref{eq:ns5a} vanishes. For $\hat{n} > 0$, such a change of variables yields the following non-local reaction equation
		\begin{equation}
		\label{eq:ns5}
		\begin{cases}
		\partial_t \hat{u} =  \alpha \left(r(y)\hat{S} + f(y) - \hat{\rho} \right) \, , \quad (\boldsymbol{x},y) \in \overline{\Omega} \times [0,1] 
		\\\\
	\displaystyle{\hat{\rho}(t,\boldsymbol{x}) := \int_{0}^1 e^{\hat{u}(t,\boldsymbol{x},y)} \,  {\rm d}y \ ,}
		\end{cases}
	\end{equation}
which is easier to be numerically solved implicitly than the non-local reaction equation~\eqref{eq:ns5a}. 

\subsubsection{Summary of the scheme}
We consider a uniform discretisation of the phenotype domain $[0,1]$ of step $\Delta y := 1/N_y$, with $N_y \in \mathbb{N}$, whereby $y_j = j \ \Delta y$ for $j=0, \ldots, N_y$. Similarly, we consider a uniform discretisation of the time domain $[0,t_f]$ of step $\Delta t := t_f/N_t$, with $t_f \in \mathbb{R}^+$ and $N_t \in \mathbb{N}$, whereby $t_k = k \ \Delta t$ for $k=0, \ldots, N_t$. On the other hand, we employ a finite-element discretisation for the spatial domain $\Omega$ by considering the following finite element spaces
$$
V_h := \{f \in C^0(\overline{\Omega}) \, : \, f|_T \in P_1(T) \; \forall T \in \mathcal{T}_h \} \subset H^1(\Omega) \ ,
$$	
$$
    V_h^0 := \left\{ f \in V_h \; : \; f \equiv 0 \text{ on } \partial\Omega \right\} \subset H_0^1(\Omega) \ ,
$$
$$
    V_h^1 := \left\{ f \in V_h \; : \; f \equiv 1 \text{ on } \partial\Omega \right\} \ ,
$$
where $\mathcal{T}_h$ is a partition of $\Omega$ into triangular ($d = 2$) or tetrahedral ($d = 3$) elements and $P_1(T)$ is the space of polynomials of order $1$ on the tetrahedral element $T$. Here, the numerical approximations of $\hat{S}(t_k, \boldsymbol{x}) \in V_h^1$, $\hat{n}(t_k, \boldsymbol{x}, y_j) \in V_h$, and $\hat{u}(t_k, \boldsymbol{x}, y_j) \in V_h$ are denoted by $S^k$, $n^k_j$, and $u^k_j$, respectively, and the standard $L^2$-inner product over $\Omega$ is denoted by $<\cdot, \cdot>$. 

First we solve the diffusion equation~\eqref{eq:ns1} subject to the boundary conditions~\eqref{eq:BCsS} through the following finite element scheme
	\begin{equation}
		\left<\dfrac{{S}^* - S^k}{\Delta t}, v_h\right> = - \left<{\bf D}_S(\boldsymbol{x}) \,  \nabla S^* , \nabla v_h\right> \, , \quad \forall v_h\in V_h^0
	\end{equation}
to obtain ${S}^*$, and subsequently we compute $S^{k+1}$ by solving the reaction equation~\eqref{eq:ns2} through the following finite difference scheme 
	\begin{equation}
		\dfrac{S^{k+1} - {S}^*}{\Delta t} = - \hat{\nu} \ S^{k+1} \ \sum_{j=1}^{N_y} \frac{r_{j-1} n_{j-1}^{k}+ r_j n_j^{k}}{2} \ \Delta y \, ,
	\end{equation}
	where $r_j = r(y_j)$.
	
Next we solve the diffusion equation~\eqref{eq:ns3} subject to the boundary conditions~\eqref{eq:BCsnx} through the following finite element scheme
	\begin{equation}
		\left<\dfrac{n_j^{*}-n_j^k}{\Delta t}, v_h\right> = - \left< \mathbf{D}_n(\boldsymbol{x}) \, \nabla n^*_j,\nabla{v}_h \right> \, , \quad \forall v_h\in V_h, \quad j=0, \ldots, N_y
	\end{equation}
to obtain $n_j^{*}$. 

Then, looping over all the elements of $\mathcal{T}_h$, we solve the diffusion equation~\eqref{eq:ns4} subject to the boundary conditions~\eqref{eq:BCsny} through the following finite difference scheme 
	\begin{equation}
	\begin{cases}
	\displaystyle{\frac{n_j^{**} - n_j^{*}}{\Delta t} = \beta \frac{n_{j+1}^{**} + n_{j-1}^{**} - 2 n_j^{**}}{(\Delta y)^2}, \quad j=0, \ldots, N_y \; }
		\\
			\displaystyle{n_{-1}^{**} = n_{1}^{**}, \quad n_{N_y+1}^{**} = n_{N_y-1}^{**} }
	\end{cases}
	\end{equation}
to obtain $n_j^{**}$, where the ghost points $j=-1$ and $j=N_y+1$ have been introduced to gain second-order numerical accuracy. 

Finally, employing the following finite difference scheme for the non-local reaction equation~\eqref{eq:ns5} 
		\begin{equation}
		\label{scheme:u}
\displaystyle{\frac{u_j^{k+1} - u_j^{**}}{\Delta t} = \alpha \left( r_j S^{k+1} + f_j -  \rho^{k+1} \right) \, , \quad  j=0, \ldots, N_y \, },
	\end{equation}
and recalling that $u_j^{k+1} = \ln(n_j^{k+1})$ and $u_j^{**} = \ln(n_j^{**})$, we compute 
\begin{equation}
\label{scheme:n}
n_j^{k+1} = n_j^{**} e^{\left[ \alpha \Delta t \left( r_j S^{k+1} + f_j -  \rho^{k+1} \right) \right]} \, , \quad  j=0, \ldots, N_y \, ,
\end{equation}
where $r_j = r(y_j)$ and $f_j = f(y_j)$, and $\rho^{k+1}$ is obtained by solving, through a root-finding algorithm, the following transcendental equation
	\begin{equation}
	\begin{aligned}
		\label{scheme:rho}
		\rho^{k+1} &= \sum_{j=1}^{N_y} \frac{n_{j-1}^{k+1} + n_j^{k+1}}{2} \Delta y  \\ &=\Delta y \,  e^{\left(- \alpha \Delta t \rho^{k+1}\right)}\sum_{j=1}^{N_y} \frac{m_{j-1}^{k+1} + m_j^{k+1}}{2} \; \text{ with } \; m_j^{k+1} = n_j^{**} e^{\left[\alpha \Delta t\left(r_j S^{k+1} + f_j\right)\right]} \ .
		\end{aligned}
	\end{equation}
The above transcendental equation is derived, with a little algebra, by first approximating the integral in the definition~\eqref{eq:ns5a} of $\hat{\rho}$ with the corresponding Riemann sum and then using~\eqref{scheme:n}. 

Note that, in the implementation of the numerical scheme, the update is performed directly through~\eqref{scheme:n} and not through~\eqref{scheme:u}, in order to avoid having to compute the logarithms. Note also that~\eqref{scheme:n} is well-defined even at points where $n_j^{**}=0$, for which it yields $n_j^{k+1}=0$, consistently with the fact that, as mentioned earlier, where $\hat{n} = 0$ the right-hand side of the non-local reaction equation~\eqref{eq:ns5a} vanishes.

	\subsection{Simulations of the rescaled one-dimensional model~\eqref{rescaled_equation_2D}}
	\label{Rescaled_model}
	\subsubsection{Set-up of simulations}
	\label{Rescaled_model:setup}
	We solve numerically the initial-value problem defined by the rescaled RD system~\eqref{rescaled_equation_2D}, with $x \in [0, L]$ and $L=3 \times 10^4$, subject to the boundary conditions~\eqref{eq:BCsnx}-\eqref{eq:BCsny} for the cell density, and Dirichlet boundary conditions $S_{\varepsilon}(t,0) = 0$ and $S_{\varepsilon}(t,L) = 1$ for the nutrient concentration, together with the following initial data:
	\begin{equation}
		\label{initial_condition}
		\begin{split}
			n_{\varepsilon}(0, x, y)= \dfrac{e^{-\frac{(y-0.5)^2}{\varepsilon}}}{\displaystyle{\int_{0}^1 e^{-\frac{(y-0.5)^2}{\varepsilon}} \, {\rm d}y}} e^{-x^2} \, , \quad S_{\varepsilon}(0, x) = 1- \int_{0}^1 n_{\e}(0,x,y) \, {\rm d}y \, = 1 - \rho_{\e}(0, x) \, .
		\end{split}
	\end{equation}
We set $\Delta y = 10^{-2}$ and $\Delta t = 10^{-3}$. This value of $\Delta t$, which is chosen to be sufficiently small so as to accurately capture the emergence of concentration phenomena and the formation of steep travelling fronts, was selected after preliminary tests, which indicated that smaller values led to propagating fronts of overall the same shape but at a higher computational cost.

All simulations are performed considering dimensionless space and time variables, and using the definitions~\eqref{def:modparfun} of the model functions, with $\gamma = 1$ and $\zeta = 0.1$. Moreover, since the formal asymptotic results of Section~\ref{formal_asymptotic_analysis} are obtained by letting first $\e \to 0$ and then $\alpha \to \infty$, we choose $\e=10^{-4}$ and $\alpha = 20$. Note that the excellent quantitative agreement between the analytical results presented in Section~\ref{formal_asymptotic_analysis} and the numerical results presented in this section (cf. Figure~\ref{plot_rescaled}) indicates that these values of $\e$ and $\alpha$ can be regarded, respectively, as suitably small and suitably large for the employed model set-up. It is also worth noting that, since the numerical simulations we report on in this section are primarily meant to verify the analytical results obtained, the parameter values employed here are not calibrated to specific biological data, as them and their relative ratios are chosen with the main purpose of satisfying the assumptions underlying the formal analyses. In addition, the spatial domain was chosen to be sufficiently large to allow for the full formation of travelling fronts.
	
\subsubsection{Main results}	
\begin{figure}
		\centering
		\includegraphics[width=0.8\textwidth]{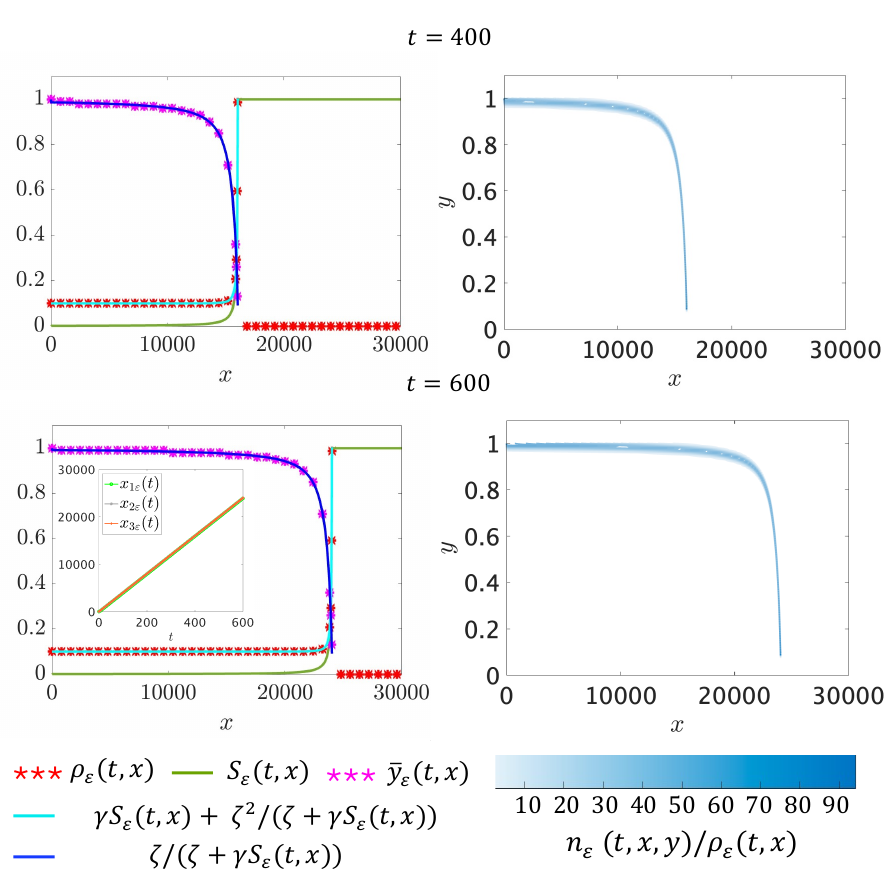}
\caption{{\bf Simulations of the rescaled one-dimensional model~\eqref{rescaled_equation_2D}.} {\bf Left panels.} Plots of $\rho_{\e}(t,x)$ (red), $S_{\e}(t,x)$ (green), and $\bar{y}_{\e}(t,x)$ (magenta), with $\bar{y}_{\e}(t,x)$ being the maximum point of $n_{\e}/\rho_{\e}$ along the $y$-dimension at $x \in {\rm Supp}(\rho_{\e})$, for $t=400$ (top panel) and $t=600$ (bottom panel). The blue line and the cyan line highlight, respectively, the predicted analytical values of $\bar{y}_{\e}$ and $\rho_{\e}$, which are obtained by inserting the numerical values of $S_{\e}$ into the formulas given by~\eqref{bar_yspec2} and~\eqref{rhospec2}. The inset of the bottom panel displays the plots of $x_{1\e}(t)$ (green), $x_{2\e}(t)$ (grey), and $x_{3\e}(t)$ (orange), defined by the level sets $\rho_{\e}(t,x_{1\e}(t))=0.3$, $\rho_{\e}(t,x_{2\e}(t))=0.5$, and $\rho_{\e}(t,x_{3\e}(t))=0.7$, respectively. These are straight lines overlapping almost perfectly, whose common slope, which is approximately $40$, provides a numerical estimate of the speed of the travelling front.
{\bf Right panels.} Plots of $n_{\e}/\rho_{\e}$ for $x \in {\rm Supp}(\rho_{\e})$ at $t=400$ (top panel) and $t=600$ (bottom panel), with the corresponding colour scale displayed below the bottom panel. The set-up of simulations is summarised in Section~\ref{Rescaled_model:setup}.}
		\label{plot_rescaled}
	\end{figure}

The right panels of Figure~\ref{plot_rescaled} display the plots of the normalised local phenotype density, $n_{\e}/\rho_{\e}$, for $x \in {\rm Supp}(\rho_{\e})$, at two successive time instants, i.e. $t=400$ (top panel) and $t=600$ (bottom panel). These plots indicate that, for all $x \in {\rm Supp}(\rho_{\e})$, the normalised local phenotype density is concentrated as a sharp Gaussian with maximum at a point, $\bar{y}_{\e}(t,x)$, which can be regarded as the locally prevailing cell phenotype -- i.e. $n_{\e}(t,x,y)/\rho_{\e}(t,x) \approx \delta_{\bar{y}_{\e}(t,x)}(y)$ for all $x \in {\rm Supp}(\rho_{\e})$.

As shown by the plots in the left panels of Figure~\ref{plot_rescaled}, the locally prevailing cell phenotype (i.e. the maximum point of $n_{\e}/\rho_{\e}$ along the $y$-dimension at $x \in {\rm Supp}(\rho_{\e})$), $\bar{y}_{\e}(t,x)$,  the cell density, $\rho_{\e}(t,x)$, and the oxygen concentration, $S_{\e}(t,x)$, behave like monotonic travelling fronts. In particular, on ${\rm Supp}(\rho_{\e})$, the numerical values of $\bar{y}_{\e}$ and $\rho_{\e}$ coincide with the predicted analytical values obtained by inserting the numerical values of $S_{\e}$ into the formulas given by~\eqref{bar_yspec2} and~\eqref{rhospec2}, respectively. Consistently with the results of the formal asymptotic analyses presented in Section~\ref{formal_asymptotic_analysis}, the value of $S_{\e}$ grows monotonically from $0$, attained in the core of the tumour, to $1$, attained ahead of the tumour. Furthermore, $\rho_{\e}$ undergoes a steep transition to zero at the tumour border. In the limit $\varepsilon \to 0$, this transition becomes the jump predicted by \eqref{eq:shapeofrho} and \eqref{rho_bary_edge}. For $\varepsilon > 0$, it is instead continuous and takes place within a thin layer, the width of which decreases as the value of $\varepsilon$ decreases. For the small value of $\varepsilon$ considered here, this layer is not fully captured by the spatial discretisation, and the transition therefore appears as a jump in Figure~\ref{plot_rescaled}.


\noindent To estimate the speed of the travelling front, we track the positions $x_{1\e}(t)$, $x_{2\e}(t)$, and $x_{3\e}(t)$ defined by the level sets $\rho_{\e}(t,x_{1\e}(t))=0.3$, $\rho_{\e}(t,x_{2\e}(t))~=~0.5$, and $\rho_{\e}(t,x_{3\e}(t))~=~0.7$, respectively. The plots in the inset of the left-bottom panel of Figure~\ref{plot_rescaled} show that $x_{1\e}(t)$, $x_{2\e}(t)$, and $x_{3\e}(t)$ are straight lines overlapping almost perfectly, which indicates that $\rho_\varepsilon$ behaves like a travelling front. The common slope of these lines provides a numerical estimate of the wave speed, which is approximately $40$ (in dimensionless units). This is consistent with the analytically predicted value computed through~\eqref{wave_speed}, which is approximately $38.9$ (in dimensionless units). To compute this analytically predicted value, we approximated the integral in~\eqref{wave_speed} by a Riemann sum and then substituted the numerical values of $\bar{y}_{\e}(t,x)$ and $S_{\e}(t,x)$ into the resulting formula, with $r(y)$ and $f(y)$ defined via~\eqref{def:modparfun}. We remark that an equivalent tracking of $x_{4\e}(t)$, $x_{5\e}(t)$, and $x_{6\e}(t)$ defined by the level sets $\bar{y}_{\e}(t,x_{4\e}(t))=0.2$, $\bar{y}_{\e}(t,x_{5\e}(t))~=~0.4$, and $\bar{y}_{\e}(t,x_{6\e}(t))~=~0.6$, respectively, yields quasi-identical results.

\newpage
	\subsection{Simulations of the model~\eqref{modello} posed on a square spatial domain}
	\label{Non_rescaled_model_in_a_square_domain}
	\subsubsection{Set-up of simulations}
		\label{Non_rescaled_model_in_a_square_domain:setup}		
	We pose the RD system~\eqref{modello} on the square spatial domain $[-L,L] \times [-L,L]$ with $L=75 \ \rm mm$. 
	We complement the system~\eqref{modello} with the boundary conditions~\eqref{eq:BCsnx}-\eqref{eq:BCsS} and the following initial data 
	\begin{equation}
		\label{initial_condition_non_rescaled}
		\begin{split}
			\hat{n}(0, \boldsymbol{x}, y)= \dfrac{e^{-\frac{(y-0.2)^2}{0.1}}}{\displaystyle{\int_{0}^1 e^{-\frac{(y-0.2)^2}{0.1}} \, {\rm d}y}} e^{-|\boldsymbol{x}|^2} \, , \quad \hat{S}(0, \boldsymbol{x}) = 1- \int_{0}^1 \hat{n}(0,\boldsymbol{x},y) \, {\rm d}y \, = 1 - \hat{\rho}(0, \boldsymbol{x}) \, .
		\end{split}
	\end{equation}
We set $\Delta y = 10^{-2}$ and $\Delta t = 0.1$. Such a value of $\Delta t$ ensures an appropriate compromise between numerical robustness and computational efficiency, as shown by preliminary refinement tests which confirmed that this value led to stable and qualitatively consistent propagation dynamics. Note that, in contrast to the numerical simulations of Section~\ref{Rescaled_model}, here numerical simulations are carried out for the original model instead of the $\varepsilon-$scaled one and, therefore, it is to be expected that a less refined temporal discretisation is required.
	
We choose $\alpha=1 \ \mathrm{day}^{-1}$. This corresponds to a normalisation of the time scale, which is without loss of generality, as the scaling effect can be directly absorbed into the parameters $\gamma$, $\zeta$, and $\rho_0$. In particular, under this choice of the value of $\alpha$, the parameters $\gamma$ and $\zeta$ can be regarded as the maximum proliferation rates via aerobic respiration and anaerobic glycolysis, respectively. Moreover, we use the definitions~\eqref{def:modparfun} of the model functions along with
	\begin{equation}
		\label{assumption_2}
		{\bf D}_n(\boldsymbol{x}):= D_n{\bf D}(\boldsymbol{x}) \; \text{ and } \; {\bf D}_S(\boldsymbol{x}):= D_S{\bf D}(\boldsymbol{x}) \ ,
	\end{equation}
with ${\bf D}(\boldsymbol{x}):={\rm diag}(D_1,D_2)$. Specifically, to investigate the impact of anisotropy in cell movement and oxygen diffusion on the dynamics of tumour cells, we set $D_1=1.6$ and $D_2=0.4$. These values of $D_1$ and $D_2$ are chosen to enhance horizontal migration along the first component of $\boldsymbol{x}$ (i.e. $D_1 > D_2$) while preserving the trace of the diffusion tensor, and thus the mean diffusivity, which is given by the trace divided by the spatial dimension. This ensures that only the anisotropy of the process is modified, without altering the overall diffusion strength. The values of the other model parameters are those listed in Table~\ref{tabella_parametri}, which are chosen to be consistent with the existing literature (cf. the references provided in Table~\ref{tabella_parametri}). We note that, taken together, these parameter values correspond to a fast-growing and highly aggressive tumour, a choice we make to facilitate early emergence of observable effects of phenotypic evolution dynamics, in order to reduce the computational time of simulations. However, we remark that the phenotype-structured RD system~\eqref{modello} constitutes a flexible modelling framework that can also accommodate parameter values for slow-growing and less aggressive tumours.

	\begin{table}
		\small
		\centering
		\begin{tabular}{|c|l|l|l|l|} 
			\hline
			Par. & Biological meaning & Value & Units & Refs. \\
			\hline
			$\rho_0$ & Local carrying capacity of the tumour & $3.183 \times 10^{3}$ & $\mathrm{cells}/\mathrm{mm}^2$ & \cite{Gu:2011} \\
			$S_0$ & Tissue oxygen conc. in physiological conditions & $6.3996 \times 10^{-9}$ & \ $\mathrm{g}/\mathrm{mm}^2$ & \cite{Kumosa:2014, Villa:2021} \\
			$\hat{\nu}$ & Rate of oxygen consumption & $9.95$ & $\mathrm{day}^{-1}$ & \cite{Lorenzi:2018, Villa:2021} \\ 
			$\gamma$ & Maximum prolif. rate via aerobic respiration & $0.864$ & -  & \cite{Gordan:2007,Lorenzi:2018} \\ 
			$\zeta$ & Maximum prolif. rate via anaerobic glycolysis & $0.0864$ &- & \cite{Gordan:2007,Lorenzi:2018} \\ 
			$\beta$ & Rate of phenotypic changes & $8.64 \times 10^{-8}$  & $\mathrm{day}^{-1}$ & \cite{Fiandaca:2021, Villa:2021} \\
			$D_n$ & Diffusion coefficient of glioma cells & $0.13$ & $\mathrm{mm}^2/\mathrm{day}$  & \cite{Swanson:2000} \\ 
			$D_S$ & Diffusion coefficient of oxygen & $86.4$ & $\mathrm{mm}^2/\mathrm{day}$ & \cite{Hlatky:1985,Villa:2021} \\ 
			\hline
		\end{tabular}
		\caption{List of the values of the model parameters employed for the numerical simulations in Sections~\ref{Non_rescaled_model_in_a_square_domain} and~\ref{sec:brainsim}.}
		\label{tabella_parametri}
	\end{table}

\subsubsection{Main results}	
\begin{figure}
		\centering
		\includegraphics[width=0.8\textwidth]{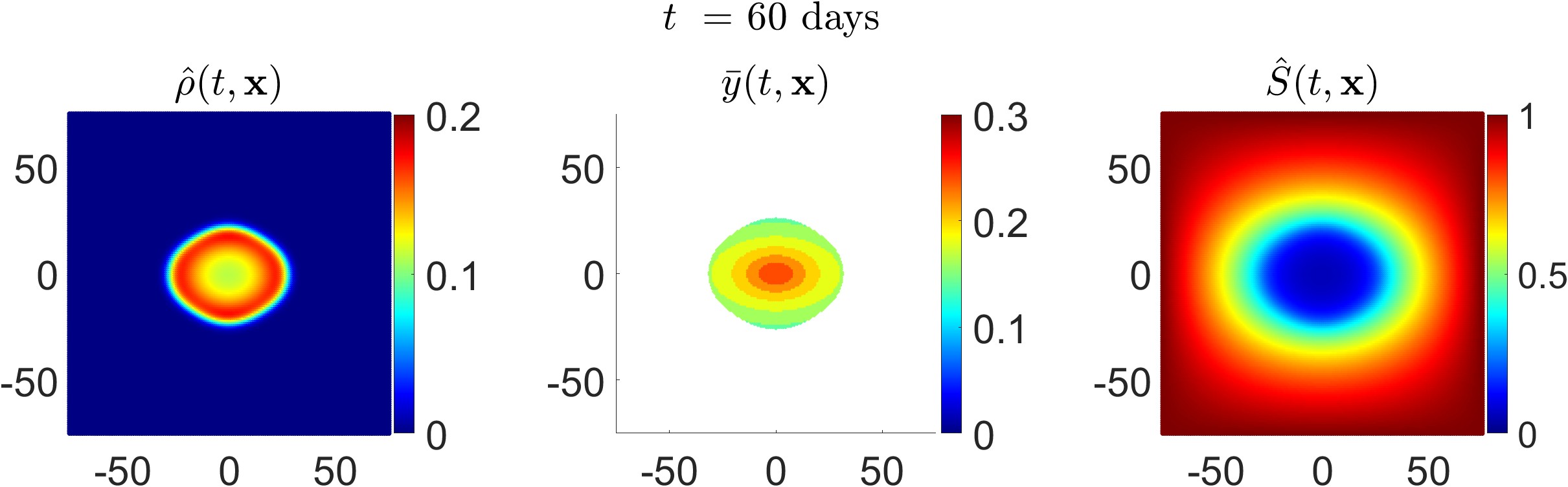} \\ \vspace{5mm}
		\includegraphics[width=0.8\textwidth]{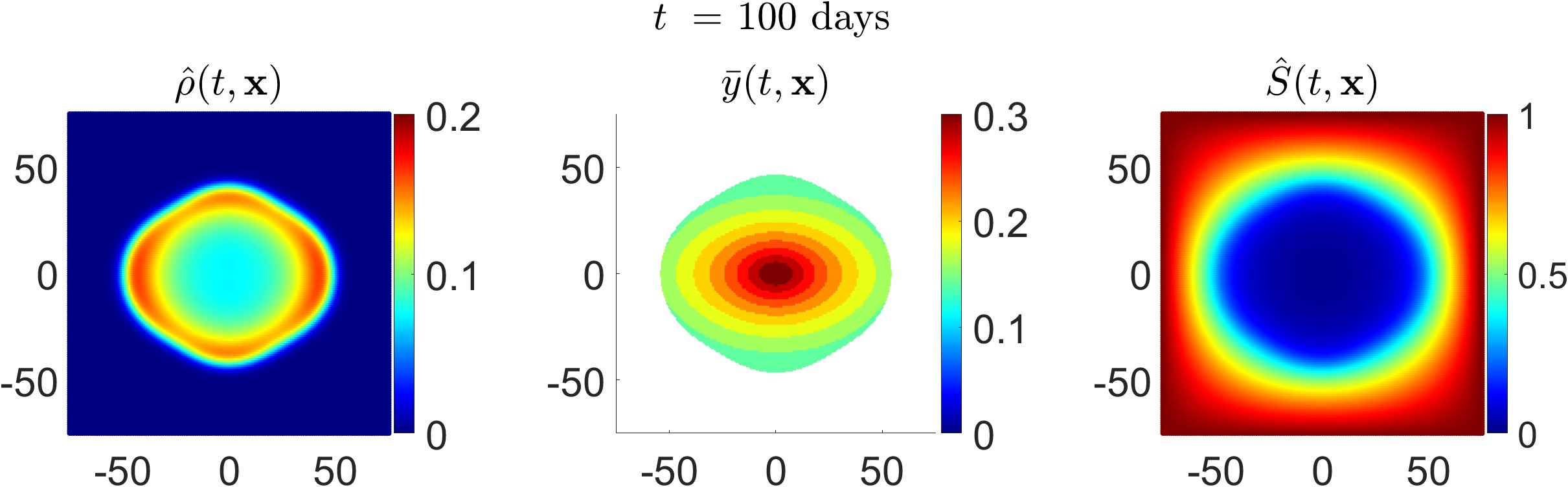} 
\caption{{\bf Simulations of the model~\eqref{modello} posed on a square spatial domain.} Plots of $\hat{\rho}(t,\boldsymbol{x})$ (left), $\bar{y}(t,\boldsymbol{x})$ defined via~\eqref{def:barynum} (centre), and $\hat{S}(t,\boldsymbol{x})$ (right), at time $t=60$ days (top panels) and $t=100$ days (bottom panels). 
}		
		\label{plot2D_non_rescaled}
	\end{figure}

The dynamics of numerical solutions are summarised by the plots in Figure~\ref{plot2D_non_rescaled}, which display the rescaled cell density, $\hat{\rho}(t,\boldsymbol{x})$, the locally prevailing cell phenotype, $\bar{y}(t,\boldsymbol{x})$, that is,
\begin{equation}
\label{def:barynum}
\hat{n}(t,\boldsymbol{x},\bar{y}(t,\boldsymbol{x})) = \max_{y \in [0,1]} \hat{n}(t,\boldsymbol{x},y) \ ,
\end{equation}
and the rescaled oxygen concentration, $\hat{S}(t, \boldsymbol{x})$, at the times $t=60$ days and $t=100$ days. 

Consistently with what observed in the results of simulations of the one-dimensional model~\eqref{rescaled_equation_2D} displayed in Figure~\ref{plot_rescaled}, the plots in Figure~\ref{plot2D_non_rescaled} show that the rescaled oxygen concentration exhibits monotonic behaviour, being approximately $0$ at the centre of the tumour, where a hypoxic core is formed, and being instead $1$ in regions that have not yet been invaded by tumour cells. Accordingly, the locally prevailing cell phenotype decreases monotonically moving from the core to the edge of the tumour, indicating that the hypoxic core is occupied by cells with a less oxidative metabolism (i.e. cells with phenotypes corresponding to larger values of $y$), which rely less on aerobic respiration to produce the energy required for cell division, and thus proliferate more slowly, while the more oxygenated regions towards the tumour edge are occupied by cells with a more oxidative metabolism (i.e. cells with phenotypes corresponding to smaller values of $y$), which rely more on aerobic respiration for producing the energy to fuel cell division, and thus proliferate more quickly. As a result of this, the rescaled cell density is also monotone, being minimal in the hypoxic core and maximal at the tumour edge. 
We remark that, unlike the simulation results of the rescaled model~\eqref{rescaled_equation_2D}, where the cell density exhibits a sharp transition to zero at the tumour border, here we observe instead a smoother transition layer. This is to be expected, since here there is no $\varepsilon$-scaling.
Furthermore, we note that, for the parameter values considered here, phenotype structuring across the tumour has not yet fully developed by the end of the simulations (i.e. the full range of the local phenotype density has not been reached), which explains why the values of the locally prevailing cell phenotype are relatively low and span a relatively narrow range throughout the tumour.

In addition, the results in Figure~\ref{plot2D_non_rescaled} demonstrate the impact of anisotropy in cell movement and oxygen diffusion on tumour growth and on the phenotypic composition of the tumour edge. Specifically, these results indicate that, compared to the other directions, along the anisotropy preferential direction both tumour growth (i.e. the expansion of the support of $\hat{\rho}$) and oxygen consumption (i.e. the decay of $\hat{S}$) are faster, and the tumour edge comprises cells with a less oxidative metabolism (i.e. the values attained by $\bar{y}$ in the proximity of the outer boundary of the tumour are larger in the anisotropy preferential direction than in the other directions).
	
	\subsection{Simulations of the model~\eqref{modello} posed on the 3D-geometry of the brain}
	\label{sec:brainsim}
	
	\subsubsection{Set-up of simulations}	
	\label{sec:brainsim:setup}
We pose the RD system~\eqref{modello} on the 3D-geometry of the brain. The corresponding computational mesh was constructed by using Magnetic Resonance Imaging (MRI) data from a single patient, which had been acquired during routine clinical practice at the Istituto Neurologico Carlo Besta in Milan, Italy. Moreover, we define the diffusion tensors ${\bf D}_n(\boldsymbol{x})$ and ${\bf D}_S(\boldsymbol{x})$ via~\eqref{assumption_2}, where the tensor ${\bf D}(\boldsymbol{x})$ is reconstructed from Diffusion Tensor Imaging (DTI) data\footnote{By capturing the anisotropic diffusion of water molecules, DTI enables the identification and visualisation of white matter tract orientations and the preferential directions of cell migration. This information is essential for modelling processes such as tumour infiltration in brain tissues. DTI provides a symmetric, positive-definite tensor that characterises water diffusivity within each voxel.}~\cite{Assaf:2008,Basser:1995} and represents the spatially varying diffusion directions. Specifically, the components derived from medical images are normalised by the mean diffusivity, so that the resulting tensor encodes only the principal diffusion directions, excluding the contribution of the mean water diffusivity.

To construct the computational mesh, we first segmented the MRI grey-scale images in order to partition them into segments and then labelled each pixel to reconstruct the brain boundary. This process was carried out by using the software package \emph{Slicer3D}~\cite{3D_Slicer}. After segmentation, we generated the computational mesh by means of \emph{Tetgen}~\cite{Tetgen}, a tool for generating tetrahedral meshes of any 3D polyhedral domain.

To construct the tensor ${\bf D}(\boldsymbol{x})$, the DTI images corresponding to the six independent components of the diffusion tensor were aligned with the MRI images through \emph{FSL} (FMRIB Software Library)~\cite{FSL}, and the values of the six components of the tensor \( \mathbf{D}(\boldsymbol{x}) \) were then defined throughout the computational mesh by means of custom scripts implemented in the \emph{VMTK} software library~\cite{vmtk}.
	
Simulations are carried out by complementing the system~\eqref{modello} with the boundary conditions~\eqref{eq:BCsnx}-\eqref{eq:BCsS} and the initial data~\eqref{initial_condition_non_rescaled}. We set $\Delta y = 10^{-2}$ and $\Delta t = 0.1$. The underlying rationale for this choice of $\Delta t$ is analogous to that provided in Section~\ref{Non_rescaled_model_in_a_square_domain}.


We choose $\alpha=1 \ \mathrm{day}^{-1}$ and use the definitions~\eqref{def:modparfun} of the model functions, and we set the values of the model parameters as in Table~\ref{tabella_parametri}.

\subsubsection{Main results}	
\begin{figure}
		\centering
		\includegraphics[width=0.65\textwidth]{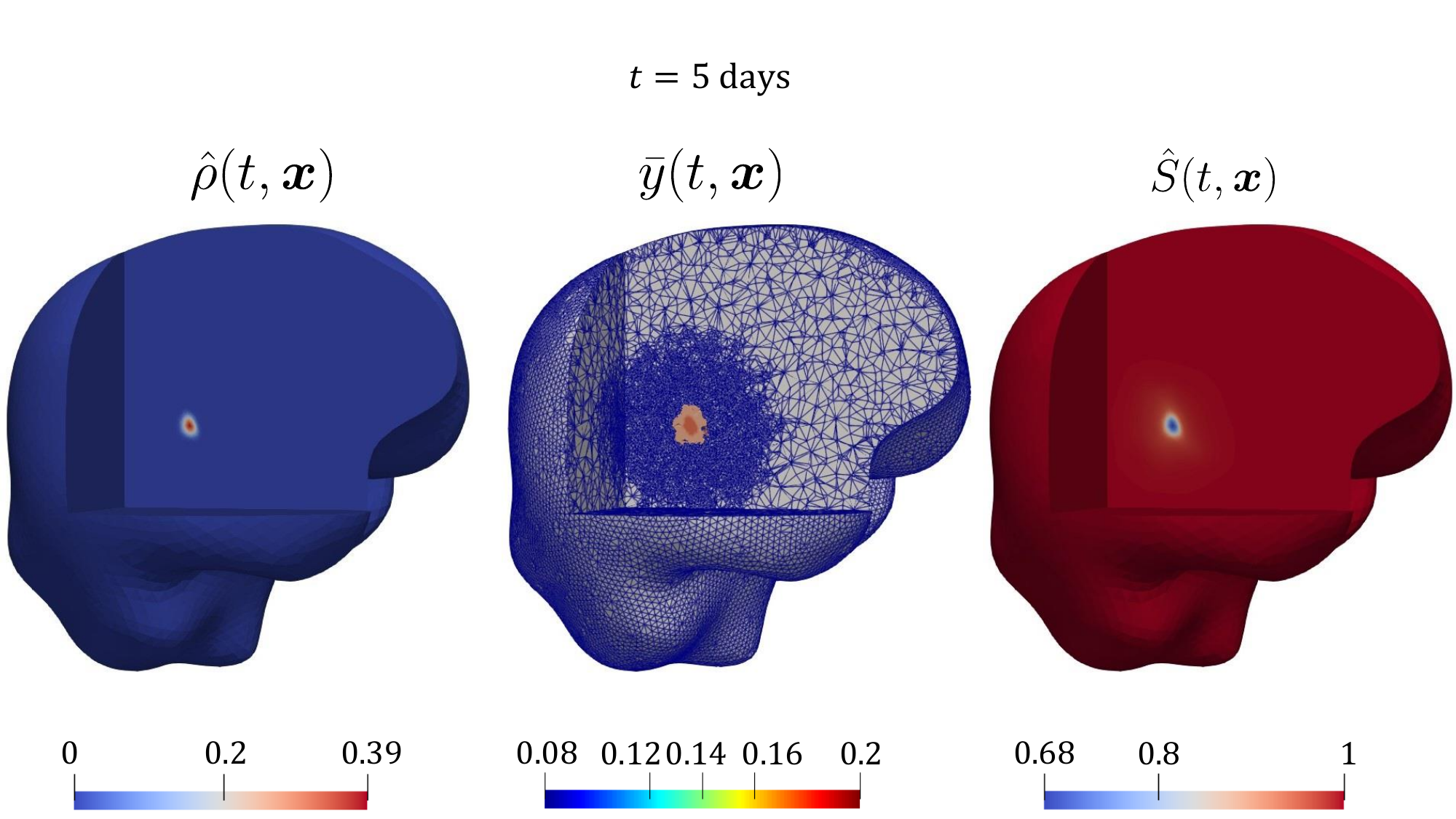}
		\includegraphics[width=0.65\textwidth]{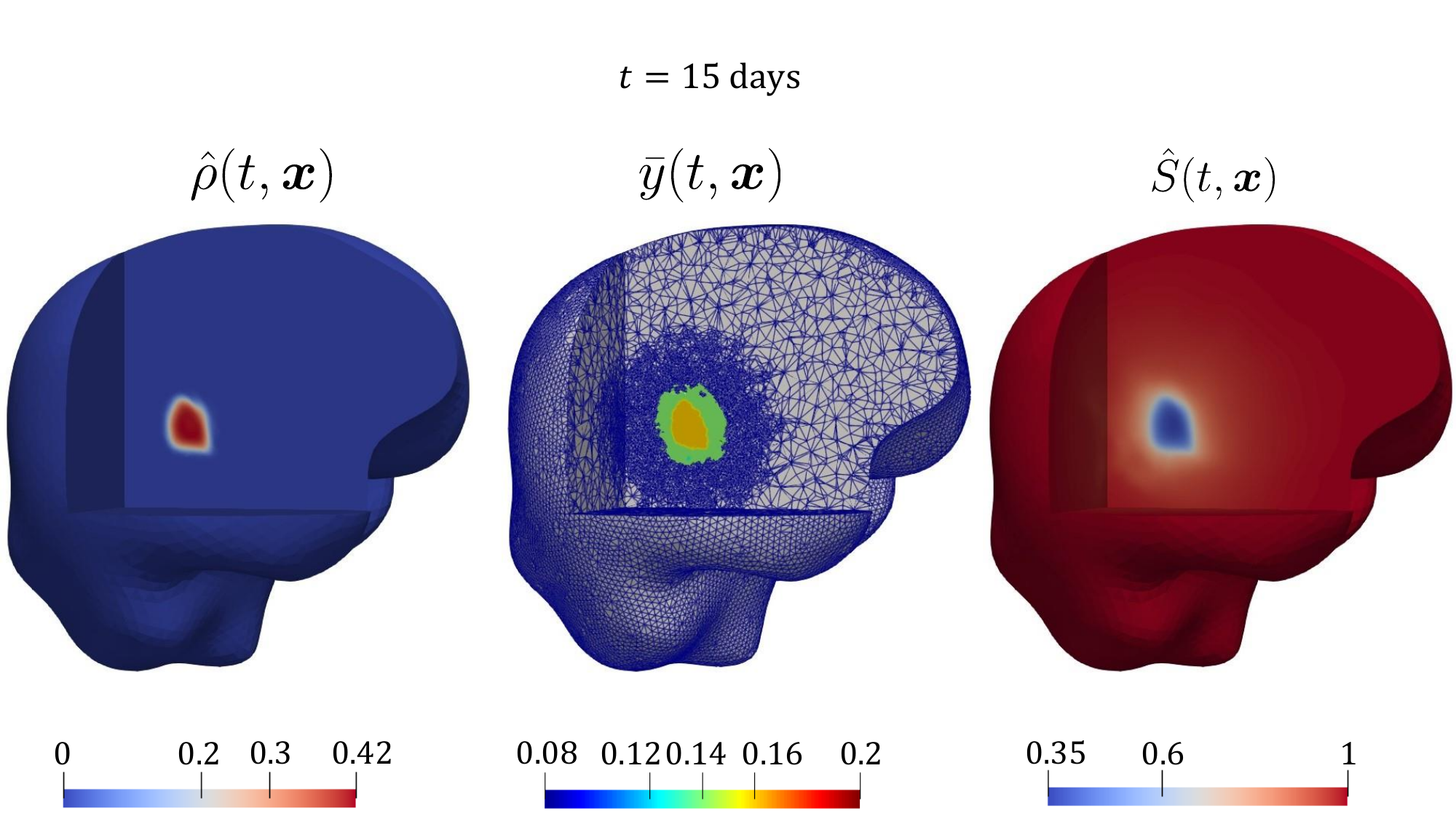}
		\includegraphics[width=0.65\textwidth]{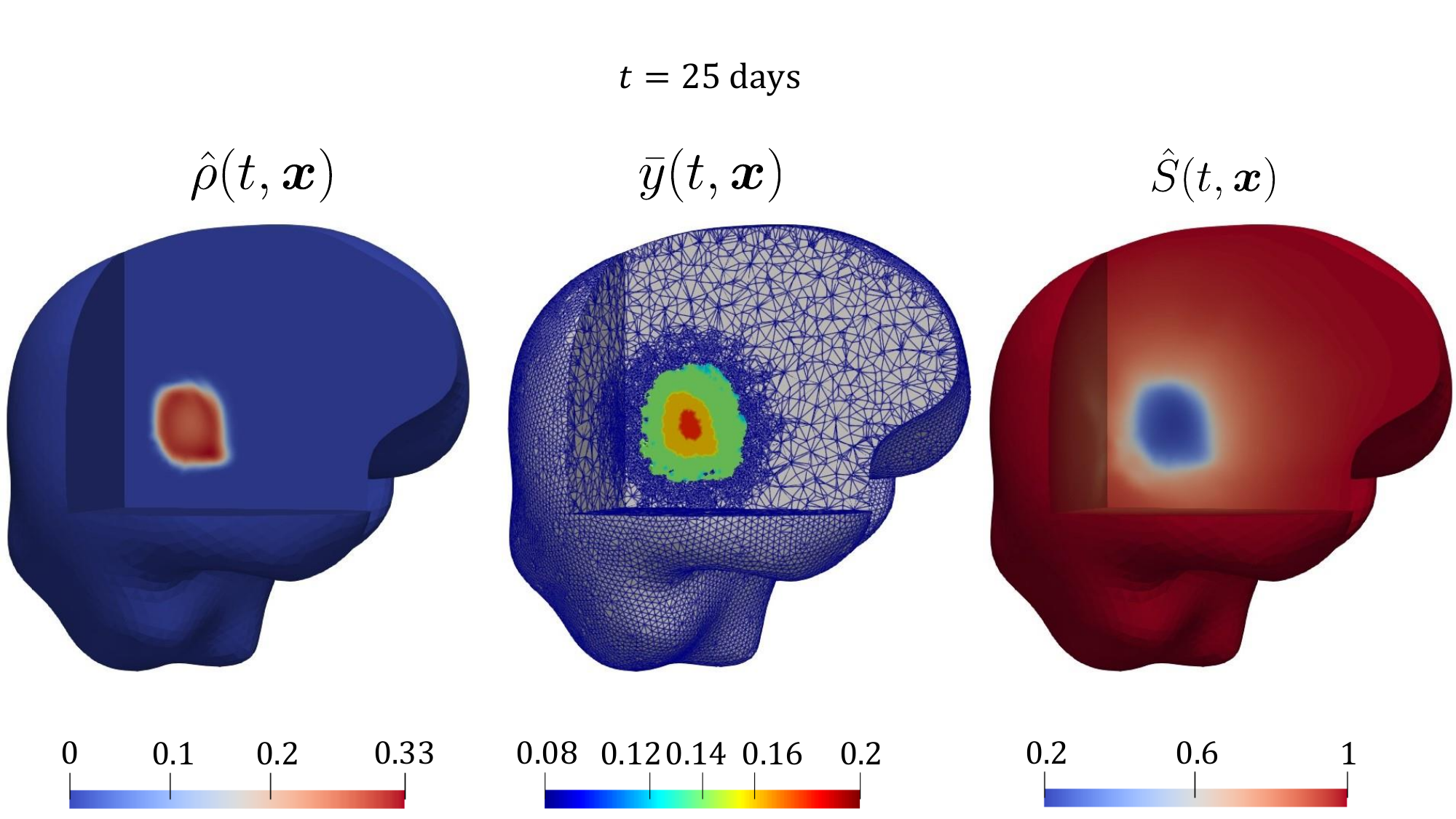}
		\includegraphics[width=0.65\textwidth]{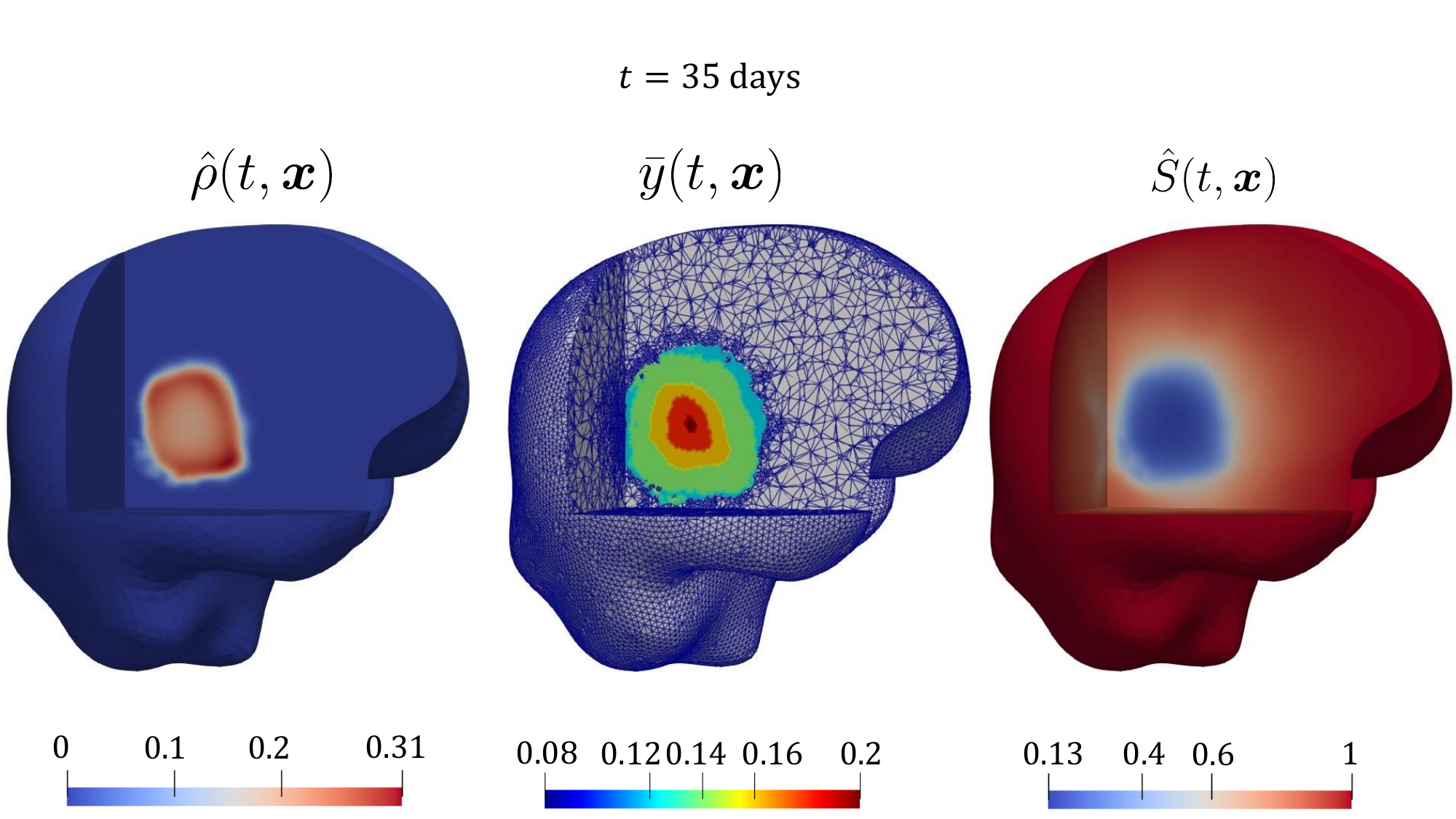}
\caption{{\bf Simulations of the model~\eqref{modello} posed on the 3D-geometry of the brain.} Plots of $\hat{\rho}(t,\boldsymbol{x})$ (left), $\bar{y}(t,\boldsymbol{x})$ defined via~\eqref{def:barynum} (centre), and $\hat{S}(t,\boldsymbol{x})$ (right), at time $t=5$ days (first row), $t=15$ days (second row), $t=25$ days (third row), and $t=35$ days (fourth row).} 
		\label{plot3D_non_rescaled}
	\end{figure}

The dynamics of numerical solutions are summarised by the plots in Figure~\ref{plot3D_non_rescaled}, which display the rescaled cell density, $\hat{\rho}(t,\boldsymbol{x})$, the locally prevailing cell phenotype, $\bar{y}(t,\boldsymbol{x})$, which is defined via~\eqref{def:barynum}, and the rescaled oxygen concentration, $\hat{S}(t, \boldsymbol{x})$, at the times $t \in \{5, 15, 25, 35\}$ days. 

These plots support the idea that the conclusions we have drawn based on the outputs of the model~\eqref{modello} posed on a square spatial domain remain intact when posing the model on the 3D-geometry of the brain (compare the plots in Figures~\ref{plot2D_non_rescaled} and~\ref{plot3D_non_rescaled}). Namely, as we move from the centre to the edge of the tumour, both the rescaled cell density and the rescaled oxygen concentration increase, while the locally prevailing cell phenotype decreases. Moreover, as the simulation time progresses, we observe the formation of a less densely populated and hypoxic tumour core, where cells exhibit a less oxidative metabolism, and a more densely populated and well oxygenated tumour edge, where cells express a more oxidative metabolism. Finally, tumour growth and oxygen consumption are faster along the preferential directions of the diffusion tensor ${\bf D}(\boldsymbol{x})$ and, in these directions of faster tumour expansion, the tumour edge is characterised by a more pronounced cell accumulation and a larger presence of cells with a less oxidative metabolism, compared to the other directions.

As mentioned earlier in the paper (cf. Section~\ref{Non_rescaled_model_in_a_square_domain:setup}), the employed parameter values listed in Table~\ref{tabella_parametri} correspond to a fast-growing and highly aggressive tumour. To verify this, we use the results of numerical simulations displayed in Figure~\ref{plot3D_non_rescaled} to compute tumour growth metrics commonly employed in the clinical literature, so as to facilitate comparison with clinical references. In summary, we estimate a volume doubling time (VDT) of approximately $5.94$ days, indicating that the tumour is growing substantially fast -- for comparison, on the basis of average data from clinical studies, Stensjøen et al.~\cite{Stensjoen:2015} reported a median VDT of $29.8$ days, while Ellingson et al.~\cite{Ellingson:2016} found a median VDT of $21.1$ days. However, highly aggressive gliomas can exhibit very high proliferation rates with doubling times ranging from 10 days~\cite{Stoyanov:2022} down to 1 week~\cite{Alvord:1991}. Thus, our simulated VDT aligns well with these worst-case clinical scenarios. The specific growth rate (SGR), or percentage increase per unit time, defined as \( (\ln 2)/\textrm{VDT} \), is $11.67\%$ per $\textrm{day}$, which is higher compared to typical values reported in~\cite{Ellingson:2016,Stensjoen:2015}, and the average radial expansion velocity (VRE) is approximately $0.62 \, \text{mm/day}$, which, as expected, exceeds average values reported in previous theoretical~\cite{Ballatore:2024} and clinical~\cite{Stensjoen:2015, Wang:2009} studies on brain tumour growth. It is also worth noting that Figure~\ref{plot3D_non_rescaled} shows that the cell density remains relatively low throughout the tumour, indicating a diffusely distributed pattern of invasion. This suggests that the simulated tumour grows predominantly by spreading its cells outwards over a broader area, rather than by increasing the concentration of cells in its immediate vicinity through rapid cell division. This behaviour is characteristic of highly invasive tumour types, in which cell dispersal dominates over local cell proliferation.


	\section{Conclusions and research perspectives}
	\label{discussion_and_research_perspectives}
The results of formal asymptotic analyses and numerical simulations presented in this work recapitulate the findings of previous experimental~\cite{Carmeliet:2000, Folkman:1971, Hanahan:2011, Sutherland:1988, Warburg:1956} and theoretical~\cite{Frieboes:2009, Gatenby:2003, Pettet:2001} studies on key features of avascular tumour growth, including the formation of a hypoxic core that comprises cells with a less oxidative metabolism, which exhibit greater ability to survive at low oxygen levels, and the appearance at the tumour's edge, where cells have access to oxygen, of a proliferative rim consisting of cells with a more oxidative metabolism. This offers a theoretical basis for experimental observations suggesting that the edge and the core of avascular tumours function as distinct ecological niches~\cite{Brown:1998, Hoefflin:2016, Lloyd:2016, Zhang:2014}, and aligns with the idea that spatial gradients of abiotic factors, such as oxygen, across the tumour play a pivotal role in the emergence of intratumour phenotypic heterogeneity~\cite{Alfarouk:2013}. 

Furthermore, the obtained results of 2D and 3D numerical simulations indicate that, along the preferential directions of cell movement and oxygen diffusion, there is faster tumour expansion and oxygen consumption, and a larger fraction of cells with less oxidative metabolism is localised at the tumour edge. These findings support the idea that structural anisotropy of the extracellular environment can impact both on tumour growth and on phenotypic selection occurring at the invasive tumour front~\cite{Finger:2025,Marino:2023,Naylor:2022,Spinnici:2025}. It would be of interest to complement these numerical results with analytical results shedding light on the impact of anisotropic diffusion on the properties of the locally prevailing cell phenotype at the leading edge of phenotypically heterogeneous travelling waves. In this respect, while the Hopf-Cole transformation used here might still prove useful, we envisage substantial changes in the formal asymptotic methods underlying our study of phenotype-structured travelling waves to be required to tackle such an intricate problem. On a related note, in the vein of~\cite{Freingruber:2025, Lorenzi:2024, Lorenzi:2022, Lorenzi:2021}, it would be interesting to further generalise the model presented here by making the diffusion tensor in the equation for the local phenotype density dependent on the phenotype-structuring variable, and then extend the analytical and numerical results presented here by examining how trade-offs between cell proliferation and migration related to the ``go-or-grow'' hypothesis, which posits a dichotomy between proliferation and migration and was conceived following observations of glioma cell behaviour~\cite{Giese:2003,Giese:1996}, may shape the phenotypic structuring of invading waves in the presence of anisotropic diffusion. Finally, while the parameter regime~\eqref{req:scaling} corresponding to a small cell diffusion coefficient, a small rate of phenotypic changes, and stronger selection is relevant for aggressive and densely packed tumours, where cells are mechanically restricted from moving, it could also be interesting to relax the smallness assumption on the cell diffusion coefficient so as to investigate scenarios corresponding to aggressive but less densely packed tumours or tumour metastasis~\cite{oswald2017jamming}.

Our approach could also extend to other types of tumours growing in different anisotropic environments and to integrate additional patient specific imaging data, such as those acquired by means of Magnetic Resonance Spectroscopy (MRS) -- an advanced MRI technique that provides information on the concentrations of water-soluble metabolites, thus enabling detection of tumour-specific mutations and assessment of intratumoural heterogeneity. MRS is frequently employed to study metabolic alterations within tumours, as it offers valuable insight into tumour grade and aggressiveness. Integrating imaging data of this type into the model could significantly improve the precision and predictive capabilities of our approach. In addition, standard-of-care imaging techniques could be exploited to inform patient-specific initial conditions. For instance, contrast-enhanced T1-weighted MRI could be used to identify the tumour core, while T2/FLAIR images could provide information on oedema and infiltrative regions, thereby enabling an estimate of the spatial distribution of the cell density. Perfusion-weighted imaging could further provide surrogate information on local blood flow, which can be used as a proxy for nutrient and oxygen availability. Moreover, histopathological data from biopsies, when available, could be used to infer, at least at an average level, the phenotypic composition of the tumour. Data-driven quantitative calibration will be essential to accurately define initial conditions and capture biologically realistic phenotypic ranges, such as the higher hypoxia levels expected in aggressive brain tumours. This opens an avenue to investigate in silico how structural anisotropy of the extracellular environment may shape the growth and phenotypic composition of solid tumours.

Building upon the modelling approach presented here, another avenue for future research would be to explore ways of incorporating a continuous phenotype structure in mechanical models of glioma growth based on either linearly elastic~\cite{Bondiau:2008, Clatz:2005, Hogea:2008} or nonlinearly elastic~\cite{Angeli:2018, Angeli:2016, Ballatore:2023, Ballatore:2024, Ehlers:2022, Ehlers:2015, Lucci:2022} constitutive equations, which would make it possible to take into account intratumour phenotypic heterogeneity when investigating the impact of brain deformations induced by tumour expansion in disease progression and patient prognosis.
	
	\section*{Acknowledgments}
	F.B., C.G., and T.L. are members of INdAM-GNFM. FB acknowledges support from the PNRR M4C2 through the project "Made in Italy Circolare e Sostenibile (MICS)", CUP: E13C22001900001. 
	X.R. acknowledges support from the R\&D Program of Beijing Municipal Education Commission from China, grant number KM202310028016, and the National Natural Science Foundation of China, grant number 12201436. T.L. gratefully acknowledges support from the Italian Ministry of University and Research (MUR) through the grant PRIN2022-PNRR project (No. P2022Z7ZAJ) ``A Unitary Mathematical Framework for Modelling Muscular Dystrophies'' (CUP: E53D23018070001) funded by the European UnionNextGenerationEU. Computational resources were provided by HPC@PO-LITO (\url{https://hpc.polito.it/}). The neuroimaging data used in this study were kindly provided by Dr. Francesco Acerbi and Dr. Alberto Bizzi (Istituto Neurologico Carlo Besta, Milan, Italy). The authors are indebted to Aymeric Stamm and Pasquale Ciarletta for their valuable collaboration in the development of the procedures for MRI and DTI image analysis.
	
	\bibliographystyle{siamplain}
	\bibliography{bibliografia}
\end{document}